\documentclass[prd,aps,onecolumn,nofootinbib,10pt]{revtex4-2}

\usepackage{amssymb,amsfonts,amsmath,amsthm,lineno}
\usepackage{comment}
\usepackage{enumerate}
\usepackage{mathrsfs}
\usepackage{bbold}
\usepackage{cancel}
\usepackage[utf8]{inputenc}
\usepackage{xcolor}
\usepackage{slashed}
\usepackage{color}
\usepackage{empheq}
\usepackage{soul}
\usepackage{mathdots}
\usepackage{dcolumn}
\usepackage{bm}

\usepackage{physics}

\usepackage{graphicx}
\usepackage{hyperref}

\makeatletter
\newcommand{\Vast}{\bBigg@{4.75}}
\makeatother

\def\b{\beta}
\def\g{\gamma}

\def\d{\delta}

\def\e{\epsilon}

\def\f{\phi}

\def\l{\lambda}
\def\L{\Lambda}
\def\m{\mu}
\def\n{\nu}

\def\s{\sigma}

\def\th{\theta}

\def\P{\Psi}

\newcommand{\be}{\begin{equation}}
\newcommand{\ee}{\end{equation}}
\newcommand{\bea}{\begin{eqnarray}}
\newcommand{\eea}{\end{eqnarray}}

\newcommand{\CH}{\mathcal{H}}

\newcommand{\CL}{\mathcal{L}}

\newcommand{\CP}{\mathcal{P}}

\newcommand\qt\tau

\newcommand{\p}{\partial}

\renewcommand{\tilde}[1]{\widetilde{#1}}
\newcommand{\del}{\partial}

\makeatletter
\renewcommand{\@seccntformat}[1]{\csname the#1\endcsname.\,\,}
\makeatother
\allowdisplaybreaks

\let \savenumberline \numberline
\def \numberline#1{\savenumberline{#1.}}

\makeatletter
\def\@fpheader{\relax}
\makeatother

\def\[{\left[}
\def\]{\right]}
\def\({\left(}
\def\){\right)}

\def\nn{\nonumber}

\begin{document}

\title{ Quantization of Carrollian fermions}
\author{Ertu\u{g}rul Ekiz}
\email{ekize15@itu.edu.tr}
\affiliation{Department of Physics, Istanbul Technical University, 
Maslak 34469 Istanbul, T\"{u}rkiye}

\author{Emre Onur Kahya}
\email{eokahya@itu.edu.tr}
\affiliation{Department of Physics, Istanbul Technical University,
Maslak 34469 Istanbul, T\"{u}rkiye}

\author{Utku Zorba}
\email{utku.zorba@iuc.edu.tr}
\affiliation{Department Of Engineering Sciences, Istanbul University-Cerrahpaşa,
Avcilar 34320 Istanbul, T\"{u}rkiye
}

\begin{abstract}
We provide the first example of interacting quantized Carrollian Dirac fermions in four dimensions and investigate their discrete symmetries, including charge conjugation (C), parity (P), and time reversal (T) transformations. As a toy model, we couple these fermions to a Carrollian scalar field using Carrollian Yukawa theory and compute the tree-level diagram, revealing an ultralocal interaction between the Carrollian fermions and the scalar field. This interaction, widely known as a Dirac delta interaction with time-dependent factor, frequently appears in quantum physics. We then address the renormalization of the theory by employing the Wilsonian procedure at one-loop order. Furthermore, we analyze the fixed points and stability properties of Carrollian Yukawa theory, comparing them with their relativistic counterparts. Beyond the specific Yukawa model studied here, we expect that our framework will have broader applications in Carrollian physics, particularly in understanding ultralocal interactions and their role in condensed matter systems, where similar phenomena arise in strongly correlated and non-relativistic regimes.\end{abstract}

%%%%%%%%%%%%%%%%%%%%%%%%%%%%%%%%%%%%%%%%%%%%%%%%%%%%%%%%%%%%%%%%%%%%%%%%%%%%%%%%
% Document
%%%%%%%%%%%%%%%%%%%%%%%%%%%%%%%%%%%%%%%%%%%%%%%%%%%%%%%%%%%%%%%%%%%%%%%%%%%%%%%%
%%%%%%%%%%%%%%%%%%%%%%%%%%%%%%%%%%%%%%%%%%%%%%%%%%%%%%%%%%%%%%%%%%%%%%%%%%%%%%%
%%%%%%%%%%%%%%%%%%%%%%%%%%%%%%%%%%%%%%%%%%%%%%%%%%%%%%%%%%%%%%%%%%%%%%%%%%%%%%%

\maketitle

%%%%%%%%%%%%%%%%%%%%%%%%%%%
\section{Introduction}
%%%%%%%%%%%%%%%%%%%%%%%%%%%
Non-Lorentzian (NL) theories have been increasingly studied in recent years, particularly in the context of their novel applications across various domains of physics \cite{deBoer:2023fnj, Bergshoeff:2022eog, Oling:2022fft}. These symmetries arise as natural extensions of the Poincaré symmetries in two distinct limits: the $c \rightarrow \infty$ limit corresponds to Galilean symmetries \cite{Inonu:1953sp}, while the $c \rightarrow 0$ limit yields Carrollian symmetries \cite{Leblond1965, Gupta1966}. Fundamentally, the Carrollian limit corresponds to the closure of the light cone, where the structure of spacetime undergoes a significant change. As a result, the motion of particles becomes ultralocal in nature, meaning that particles with nonzero energy cannot move through space nor interact with particles at different spatial locations. This behavior is captured in the framework of ultralocal field theories, which were first introduced by Klauder in 1970 \cite{Klauder:1970cs}. For this reason, the Carrollian limit is often referred to as the ultralocal limit \cite{deBoer:2023fnj}.

Initially, Carrollian symmetries were largely confined to the realm of mathematical and theoretical studies, attracting less attention compared to Galilean symmetries. However, their increasing relevance in recent years has sparked significant interest, particularly as these structures have been found to appear naturally across various areas of physics. Specifically, Carroll symmetries have been shown to play key roles in condensed matter systems \cite{Bidussi:2021nmp, Grosvenor:2021hkn, Figueroa_O_Farrill_2023, Figueroa-OFarrill:2023qty, Bagchi:2022eui, Marsot:2022imf}, hydrodynamics \cite{deBoer:2017ing, Ciambelli:2018xat, Campoleoni:2018ltl, Petkou:2022bmz, Freidel:2022bai, Bagchi:2023ysc}, cosmology \cite{deBoer:2021jej, Avila:2023brd}, string theory \cite{Bagchi:2015nca, Bagchi:2023cfp, Bagchi:2024rje, Chen:2023esw, Blair:2023noj}, field theories \cite{Bagchi:2019xfx, Henneaux:2021yzg,  Parekh:2023xms, Banerjee:2023jpi, Koutrolikos:2023evq, Mehra:2023rmm, Kasikci:2023tvs, Kasikci:2023zdn, Ecker:2024lur, Cotler:2024xhb,Chen:2024voz, Zorba:2024jxb, Bagchi:2024ikw, Sharma:2025rug}, theories of gravity \cite{Hartong:2015xda, Bergshoeff:2017btm, Ravera:2019ize, Ali:2019jjp, Grumiller:2020elf, Gomis:2020wxp, Hansen:2021fxi, Concha:2021jnn,  Ravera:2022buz, Ekiz:2022wbi, Concha:2023bly, Concha:2024dap}, black holes \cite{Ecker:2023uwm, Aggarwal:2024yxy}, and black hole event horizons \cite{Donnay:2019jiz, Redondo_Yuste_2023}.

In 1962, Bondi, van der Burg, Metzner, and Sachs \cite{osti_4799323, Sachs:1962zza} discovered an important feature of asymptotic symmetries in asymptotically flat spacetimes. They expected that, in the limit where gravity vanishes, these asymptotic symmetries would reduce to the familiar Poincaré symmetries. However, they found that the resulting symmetry algebra was in fact an infinite-dimensional extension of the Poincaré symmetries. These symmetries are now known as BMS symmetries. About a decade ago, Duval et al. \cite{Duval:2014uoa, Duval:2014uva, Duval:2014lpa} demonstrated that the BMS symmetries in four dimensions are isomorphic to the conformal Carroll symmetries in three dimensions. More recently, it was shown that, in general, all null hypersurfaces exhibit a Carrollian structure \cite{Ciambelli:2018wre, Figueroa-OFarrill:2021sxz, Herfray:2021qmp, Mittal:2022ywl, Campoleoni:2023fug}. Since the asymptotic symmetries of flat spacetimes exhibit Carrollian symmetries, these structures have increasingly been employed in studies of flat space holography in recent years; see \cite{Pasterski:2016qvg, Donnay:2022aba, Bagchi:2022emh, Donnay:2022wvx, Raclariu:2021zjz, Pasterski:2021rjz, Nguyen:2023vfz, Nguyen:2023miw, Have:2024dff, Liu:2024llk, Ruzziconi:2024kzo, Saha:2023abr, Saha:2023hsl} and references therein\,. These developments have opened new avenues for reexamining both these geometries and the field theories defined on these hypersurfaces within the Carrollian framework \cite{Hao:2021urq}. Consequently, it is natural to explore the classical and quantum properties of theories that respect Carrollian symmetries, as highlighted by the growing body of literature summarized here.

Carrollian theories are known to be obtained in two inequivalent ways: the electric limit and the magnetic limit. The electric limit corresponds to theories involving purely time derivatives, whereas the magnetic limit corresponds to theories involving both time and spatial derivatives \cite{Henneaux:2021yzg}. In this paper, we focus on electric theories and leave the treatment of magnetic theories for future work. While the quantum aspects of Galilean fermions have been explored in various studies, including Galilean electrodynamics \cite{Banerjee:2022uqj} and Galilean Yukawa theory \cite{Sharma:2023chs}, the quantum aspects of Carrollian Dirac fermions have not been investigated in detail, aside from their classical treatment in \cite{Bergshoeff:2023vfd, Bagchi:2022eui, Koutrolikos:2023evq}.

In this paper, we initiate the quantization of Carrollian Dirac fermions and explore their discrete symmetries and renormalization using a toy model. First, we construct the quantization of the free Carrollian Dirac theory and analyze its discrete symmetries, including charge conjugation (C), parity (P), and time reversal (T) transformations. We then consider the Carrollian version of Yukawa theory as a toy model and examine the tree-level results, which lead to ultralocal interactions between fermions and scalar fields. The resulting interaction is known as a point-like interaction in quantum physics and many condensed matter systems. Our result establishes a Carrollian limit of Yukawa theory as the origin of time-dependent pointlike interactions. Building on these findings, we extend our analysis to the renormalization of the Carrollian Yukawa theory by employing the Wilsonian renormalization procedure. This procedure is applied to the study of the renormalization group (RG) flow of the coupling constant, which has been previously used for Carrollian scalars in \cite{Banerjee:2023jpi}. Finally, we provide a detailed analysis of the critical points and the stability of the fixed points in both relativistic and Carrollian Yukawa theory, shedding light on the renormalization aspects of the latter theory.

\section{Carrollian symmetries}

In this section, we summarize the Carrollian symmetries and describe the transformation of fields under Carrollian transformations as a preparation for the discussions in the subsequent sections. Considering a coordinate chart  $x^\mu = \left(ct\,, x^i\right)$\,, the Lorentz transformation law for these coordinates is expressed as $x'^\mu = \Lambda^\mu{}_{\nu}\, x^\nu\,$\,, where $\mu= 0\,,...\,,3$ and  $i=1\,,...,3$\,, denote the spacetime and spatial indices, respectively\,. We adopt the mostly minus signature for the Minkowski metric: $\eta_{\mu\nu} = diag(1\,, -1\,,...\,,-1\,)$\,.  By rescaling  $\Lambda^0{}_i = - c\, b_i$\,,and taking the limit $c\rightarrow 0$, we obtain the Carrollian transformations of the coordinates \footnote{For the Galilean limit, $\Lambda^0{}_i = - c^{-1}\, v_i$, where $v_i$ is the Galilean boost parameter\,.} 
\begin{equation}
\begin{aligned}
t' &= t - \vec{b}\,\cdot \vec{x}\,, \\
\vec{x}' &= \vec{x}\,,
\end{aligned}
\end{equation}
where $\vec{b}$ is the Carrollian boost parameter. Accordingly, Carrollian generators are obtained from their Lorentzian counterparts through specific redefinitions. The Lorentz boost generator, given by  $M_{i0} = c^{-1}\,x_i\,\partial_t - c\,t\,\partial_i$\,  is identified with the Carroll boost generator as $C_i = c\, M_{i0}$\,. Likewise, the time translation generator is redefined as $H = c\, P_0$\,, where $P_0 = c^{-1}\, \partial_t$\,. In the $c\rightarrow 0$ limit, these redefinitions lead to the Carrollian generators. As a result, the Carrollian symmetry of a four-dimensional spacetime comprises time translation $H$, space translations $P_i$, spatial rotations $J_{ij}$, and Carrollian boost generators $C_i$. The coordinate representations of these generators are given by 
\begin{equation}
H = \partial_t\,, \quad P_i = \partial_i\,, \quad J_{ij} = 2x_{[i}\,\partial_{j]}\,, \quad C_i = x_i\,\partial_t\,. 
\end{equation}
These symmetry generators obey the Carrollian algebra
\begin{equation}
\begin{aligned}
    [J_{ij}, J_{kl}] &= 4\, \delta_{[i[k}\,J_{j]l]}, \\
    [J_{ij}, P_k] &=  2\,\delta_{k[i} P_{j]}, \\
   [J_{ij}, C_k] &=  2\,\delta_{k[i} C_{j]}, \\
    [P_i, C_j] &= \delta_{ij} H\, . 
\end{aligned}
\end{equation}
The transformation of fields can similarly be deduced from the Lorentz transformation of the fields. First, let us consider a scalar field transforming under an infinitesimal Lorentz transformation as follows:
\begin{equation}
   \delta\,\phi = \omega^{\mu}{}_{\nu}\, x^\nu\, \partial_\mu\, \phi\,,
\end{equation}
where infinitesimally $\Lambda^\mu{}_{\nu} = \delta^\mu{}_{\nu} - \omega^\mu{}_\nu$\,. By identifying $\omega^{0}{}_{i} = c\, b_i\,$, and taking the $c\rightarrow 0$ limit, we obtain the Carroll transformation of the scalar field as 
\begin{equation}
   \delta\,\phi = \vec{b}\cdot\,\vec{x}\, \partial_t\, \phi\,. 
\end{equation}
Having discussed the scalar field, we now turn our attention to the Dirac field. The Carrollian transformation of the Dirac field $\Psi$ can similarly be obtained using the following rule:
\begin{equation}
   \delta\,\Psi = \omega^{\mu}{}_{\nu}\, x^\nu\, \partial_\mu\, \Psi\, - \frac{1}{4}\, \omega^{\mu\nu}\, \Gamma_{\mu\nu}\, \Psi\,,
\end{equation}
where $\Gamma_{\mu\nu}$ are the matrices in the spinor representation of the Lorentz algebra\,. Therefore, again by using the identification of $\omega^0{}_i$, we obtain the following Carrollian transformation of the Dirac fields
\begin{equation}
\delta\,\Psi = \vec{b}\cdot\,\vec{x}\, \partial_t\, \Psi\, - \frac{1}{4}\, \omega^{ij}\, \Gamma_{ij}\, \Psi\,,
\end{equation}
where we only have the rotational part of the spinor transformation since this transformation corresponds to the electric limit of the Dirac fermions (see detailed discussion regarding Carrollian Dirac fermions in \cite{Bergshoeff:2023vfd})\,. 
 In the Weyl representation, the left-handed and right-handed spinors become identical in the electric Carrollian limit.  This can be observed by examining the relativistic transformations of the Weyl spinors: 
\begin{equation}
\begin{aligned}
   \delta\,\Psi_L =& -\left (\frac{i}{2}\Theta^i\, \sigma_i +  \frac{1}{2}\omega^0{}_i\, \sigma_i\,\right)\, \Psi_L\,,  \\
    \delta\,\Psi_R =& -\left (\frac{i}{2}\Theta^i\, \sigma_i -  \frac{1}{2}\omega^0{}_i\, \sigma_i\,\right)\, \Psi_R\,,
\end{aligned}
\end{equation}
where $\Theta^i$ are the parameters of the rotation generators and $\sigma_i$ are Pauli spin matrices. Thus, in the $c\rightarrow 0$ limit, the transformation for both spinors reduces to the same form
\begin{equation}
\begin{aligned}
   \delta\,\Psi_L =& -\frac{i}{2}\Theta^i\, \sigma_i\, \Psi_L\,, \\
    \delta\,\Psi_R =& -\frac{i}{2}\Theta^i\, \sigma_i\, \Psi_R\,. 
\end{aligned}    
\end{equation}
This indicates that, in the Carrollian limit, the distinction between left-handed and right-handed spinors vanishes, and they transform identically under rotations. Further properties and implications of fermions in this limit will be explored in Section \ref{sec4}.

%%%%%%%%%%%%%%%%%%%%%%%%%%%%%%%%%%%%%%%%%%%%%%%%%%
\section{Review of Carrollian scalar field quantization} 

In this section, we will summarize the quantization of scalar field theory, building upon the results presented primarily in \cite{deBoer:2021jej,deBoer:2023fnj,Banerjee:2023jpi,Cotler:2024xhb}. Our focus will be on understanding both the free and interacting scalar field theories within the Carrollian framework.

\subsection{Free Theory}  The free scalar field theory can be derived by taking the electric Carrollian limit of the relativistic Klein-Gordon theory. This limiting procedure simplifies the dynamics and highlights features unique to the Carrollian regime. Namely, theory becomes severely ultralocal. The Klein-Gordon Lagrangian in the relativistic setting provides a natural starting point for this construction, serving as the basis for extending the framework to interacting theories.  To begin with, let us introduce the relativistic Klein-Gordon Lagrangian 
\begin{equation}
	S=\int dt\int d^3 x \CL=\int dt\int d^3 x \(\frac{1}{2 c^2}\dot{\tilde{\f}}^2-\frac{1}{2}(\del_i \tilde{\f})^2-\frac{\tilde{m}^2}{2}\tilde{\f}^2 \) \label{relativisticscalar}\,, 
\end{equation}
where $\tilde{\phi}$ represents the relativistic scalar field\,, $\dot{\tilde{\phi}} = \partial_t\,\tilde{\phi}$\,, and $\tilde{m}$ is relativistic mass parameter as well. Here we split the spacetime indices into time and space components as  $x^\mu = \left(c\, t\,,  x^i  \right)$\, for a  $3+1$  dimensional Minkowski spacetime. In order to take the electric Carroll limit,  we must first employ the rescalings  $\tilde{\phi} = c\, \phi$\,, and $\tilde{m}=\frac{M}{c}$\,.  Then,  taking the $c \rightarrow 0$ limit gives rise to the following Carrollian theory
\begin{equation}
	S_{CS}= \int dt\int d^3 x \(\frac{1}{2}\dot{{\f}}^2 -\frac{M^2}{2} {\f}^2 \)\,, \label{2}
\end{equation}
where $\phi$ is the Carrollian scalar field and $M$ is the mass parameter in the Carrollian setting.  To proceed, we introduce the field equation as 
\begin{equation}
	\ddot{\f}+M^2 \f=0\,. \label{2a}
\end{equation}  
Following this, the general solution to $\f$ can be expanded as the following general form :
\begin{equation}
	\f(t,\vec{x})= e^{i M t}\int \frac{d^3\vec{k}}{(2\pi)^3} \frac{e^{-i \Vec{k}\cdot\Vec{x}} }{\sqrt{2M}}a_{\Vec{k}}^{*}+ e^{-i M t}\int \frac{d^3\vec{k}}{(2\pi)^3} \frac{e^{i \Vec{k}\cdot\Vec{x}}}{\sqrt{2M}}a_{\Vec{k}  }\,, \label{3}    
\end{equation}
where $a_{\Vec{k}}$  and $a_{\Vec{k}}^{*}$ are regarded as integration constants at this juncture; however, we will interpret them as annihilation and creation operators in a moment, consistent with standard conventions. Therefore, the canonical quantization of the scalar field can be achieved by imposing the canonical commutation relation such as
\begin{equation}
\begin{aligned}
		\comm{\f(t,\vec{x})}{\pi(t,\vec{y})}&=i \delta^3(\Vec{x}-\Vec{y})\,,
\end{aligned}\label{5}
\end{equation}
where canonical momenta is given by the time derivative of the field operator, $\pi(t,\vec{x}) = \dot{\phi}(t\,, \vec{x})$\,. By imposing this quantization condition, the integration constants $a_{\Vec{k}}$ and $a^*_{\Vec{k}}$ becomes operators which span the Hilbert space of the scalar field theory as annihilation and creation operators, respectively \cite{deBoer:2023fnj}\,. Thus, they obey the following commutation relation
\begin{equation}
	\comm{a_{\vec{k}}}{a_{\vec{l}}^{\dag}}=(2\pi)^3\delta^3(\vec{k}-\vec{l})\,. \label{8}
\end{equation}

\subsection{Interacting Theory} To explore the interacting theory, we will introduce the $\lambda \phi^4$ interaction\,. The results for this theory, particularly at the perturbative level, have been extensively studied and detailed in \cite{Banerjee:2023jpi}.  Upon introducing the interaction term $\tilde{\l} \f^4$ into the \eqref{relativisticscalar} and suitably scaling $\tilde{\l}=\l /c^4$, we arrive at the following Carrollian self-interacting scalar field theory:
\begin{equation}
	S_{ICS}= \int dt\int d^3 x \(\frac{1}{2}\dot{{\f}}^2 -\frac{M^2}{2} {\f}^2 - \frac{\lambda}{4!} \,\phi^4 \)\,. \label{3a}
\end{equation}
At the perturbative level, Feynman Rules in the momentum basis are given by
\begin{equation}
\begin{aligned}
		\text{scalar propagator:}\qquad &  S_{SF}(w)&=&\frac{i}{w^2-M^2+i\e}\\
		\text{4-point vertex:}\qquad & V_\lambda  &=&  -i \l
\end{aligned} \label{scalarfeynmanrules}
\end{equation}
The renormalization of the $\lambda \phi^4$ theory within the Carrollian context reveals new insights into the behavior of scalar fields under this limiting process. The details about how the beta functions are derived for Carrollian scalar field theory with $\lambda\, \phi^4$ interaction again can be consultant in \cite{Banerjee:2023jpi}.

In the subsequent section, we will extend the discussion by incorporating the Carrollian Dirac field and introducing an interaction between fermion and scalar field utilizing the Yukawa theory as a toy model. This addition modifies the renormalization structure of the scalar theory, leading to slightly different beta functions, the details of which will be provided in the sequel. Notably, in the absence of the Yukawa interaction term, the beta function of the scalar theory reduces to the form established in the pure $\lambda \phi^4$ case. This interplay between scalar and fermionic fields demonstrates the subtle influence of the Carrollian limit on interacting theories.

\section{Carrollian Dirac Lagrangian} \label{sec4}
%%%%%%%%%%%%%%%%%%%%%%%%%%%%%%%%%%%%%%%%%%%%%%%%%%
In this section, we initiate a comprehensive investigation into the Carrollian limit of the relativistic Dirac theory, subsequently progressing to its quantization. The classical treatment of Carrollian Dirac fermions, including both electric and magnetic theories, was obtained in \cite{Bergshoeff:2023vfd}.  We will just consider the electric limit as a starting point in this work. After revealing the classical results, we then pave the way to the quantization of the Carrollian fermionic field.  Following this, we explore the relevant operations such as charge conjugation, parity, and time reversal in the Carrollian Dirac field. To begin with,  we consider the following relativistic Dirac action in four dimensions:
\begin{equation}
\begin{aligned}
		S=&\int  dt \int d^3 x \bar{\Tilde{\P}} \( i\g^{\m} \partial_{\m}- M_f\) \Tilde{\P}\,, \\
		=&\int dt\int d^3 x \bar{\Tilde{\P}}\(i\frac{\g^0}{c}\dot{\tilde{\P}}+i \g^j\partial_j\Tilde{\P}- M_f\Tilde{\P}    \)\,. 
\end{aligned}
\end{equation}  
Consequently, we implement the scalings delineated in \cite{Bergshoeff:2023vfd} :
\begin{equation}
	\Tilde{\P}=\sqrt{c} \P \And M_f= \frac{m}{c}\,. 
\end{equation}  
Thereafter, when we examine the $c\rightarrow0$ limit, we obtain the electric theory as 
\begin{equation}
	S_{ C D}=\int dt\int d^3x\,  \Bar{\P}\(i\g^0\dot{\P}-m\P \)\,, \label{9}
\end{equation}
where $\Bar{\P}=\P^\dag\g^0$ and the Gamma matrix components are given by 
\begin{equation}
\g^0=\begin{pmatrix} 0 & \mathbb{1}_{2\times2} \\
	\mathbb{1}_{2\times2} & 0
\end{pmatrix}\,, \qquad \g^i=\begin{pmatrix} 0 & \sigma^i  \\
	- \sigma^i  & 0
\end{pmatrix}\,,
\end{equation}
with $\mathbb{1}_{2\times2}$ denoting the $2\times 2$  identity matrix.  As usual in the context of the Carrollian limit, electric theories involve merely time derivatives even if we have an integration measure over all the coordinates. The freedom of having just time derivatives without spatial ones ultimately changes the structure of the theory (see \cite{Cotler:2024xhb}\,for a comprehensive discussion on this). Since we now have the desired Lagrangian, we can continue with its equations of motion as follows :   
\begin{equation}
	i\g^0\dot{\P}=m\P\,, \label{10}
\end{equation}
then, the general solution to this equation  is given by
\begin{equation}
	\P(t,\Vec{x})=\int \frac{d\Tilde{k}}{\sqrt{2m}}\(e^{-i m t} e^{i\Vec{k}\cdot\Vec{x}}a_{\Vec{k}}\,u+ e^{i m t} e^{-i\Vec{k}\cdot\Vec{x}}b^{*}_{\Vec{k}}\,v \)\,,  \label{20} 
\end{equation}
where $a_{\Vec{k}}$ and $b^{*}_{\Vec{k}}$ are integration constants, as usual. Note that we also adopted the notation $d\tilde{k}$ which stands for $\frac{d^3\vec{k}}{(2\pi)^3}$. In this context, $u$ signifies the positive frequency solutions, while $v$ denotes the negative frequency solutions, thereby satisfying the relationship\footnote{Note that this is identical to the relativistic solutions in the rest frame; see \cite{Peskin:1995ev}\,.}:
\begin{equation}
	\(m\g^0-m\mathbb{1}_{4\times4}\)u=0\hspace{0.1cm}\And \hspace{0.1cm}\(m\g^0 + m\mathbb{1}_{4\times4}\)v=0\,.  \label{21}
\end{equation}
These equations state that both $u$ and $v$ have two independent solutions such that
\begin{equation}
	u^s=\sqrt{m} \begin{pmatrix} \xi^s\\
		\xi^s
	\end{pmatrix},\hspace{0.1cm}v^s=\sqrt{m} \begin{pmatrix} \eta^s\\
		-\eta^s
	\end{pmatrix}\hspace{0.1cm}\text{where s=1,2}\,.\label{22}
\end{equation}
Herein, we adopt the normalization condition such that $ \xi^{r\dag} \xi^s=\delta^{r s}= \eta^{r\dag} \eta^s $ where both $\xi^s$ and $\eta^s$ are represented as two component spinors. Moreover, as we will see in a moment, the Feynman propagator of the theory necessitates the evaluation of spinor inner and outer products. Upon conducting preliminary algebraic computations by using the \eqref{22}, we obtain these products as follows:  
\begin{equation}
	\begin{aligned}
		&u^{r\dag}u^s=2m\delta^{r s}\And v^{r\dag}v^s=2m\delta^{r s}\,, \\
		&\Bar{u}^r u^s=2m \delta^{r s}\And \Bar{v}^r v^s=-2m \delta^{r s}\,,\\
		&u^s \Bar{u}^s=m\g^0+m\mathbb{1} \And v^s \Bar{v}^s=m\g^0-m\mathbb{1}\,, 
	\end{aligned} \label{24}
\end{equation}
and all other products diminish to zero. Consequently, \eqref{20} may be rewritten as:
\begin{equation}
	\P(t,\vec{x})=\int \frac{d\Tilde{k}}{\sqrt{2m}}\(e^{-i m t} e^{i\Vec{k}\cdot\Vec{x}}a^s_{\Vec{k}}u^s+ e^{i m t} e^{-i\Vec{k}\cdot\Vec{x}}b^{s*}_{\Vec{k}}v^s \)\,.  \label{23} 
\end{equation}
In this expression, the summation over $s$ is implicitly understood. The quantization of the theory can be achieved by  imposing the canonical commutation relations \footnote{The quantization of Carrollian fermions in the BMS context in two and three dimensions was studied in \cite{Yu:2022bcp, 
 Hao:2022xhq, Banerjee:2022ocj}.} (see \cite{Peskin:1995ev} for the details and conventions)
\begin{equation}
	\acomm{\P(t,\vec{x})}{\P^{\dag}(t,\vec{y})}=\delta^3(\vec{x}-\vec{y})\,, \label{CarrollDiracComm}
\end{equation}
and as a result, the integration constants become the creation and annihilation operators satisfying the following commutation relations 
\begin{equation}
	\acomm{b_{\Vec{k}}}{b^{\dag}_{\Vec{l}}} = \acomm{a_{\Vec{k}}} {a^{\dag}_{\Vec{l}}} =(2\pi)^3 \delta^3(\Vec{k}-\Vec{l})\,.  \label{19}
\end{equation}
We will denote the particles created by ${a^{\dag}_{\Vec{k}}}$ as Carrollian fermions while the particles created by ${b^{\dag}_{\Vec{k}}}$ as Carrollian anti-fermions.
Before moving on to the conserved quantities, let us obtain the on-shell energy-momentum tensor for the Carrollian Dirac field:
\begin{equation}
T^{\m\n}=\bar{\P}\,i\,\g^0\d^\m_0\p^\n\P\,. 
\end{equation}
Therefore, we can read off several components of the energy-momentum tensor as:
\begin{equation}
\CH=T^{00}=m\bar{\P}\P\,, \quad \CP^i=T^{0 i}=i\bar{\P}\g^0\p^i\P=i\P^{\dag}\p^i\P\,,
\end{equation}
where it can be seen that the $T^{i0}$ component of the energy-momentum tensor in Carrollian field theory is identically zero; see \cite{deBoer:2021jej} and \cite{Henneaux:2021yzg} for a detailed discussion. Note that the Hamiltonian is positive definite and given by 
\begin{equation}
\begin{aligned}
	H &= \int\, d^3x\,  m\, \Bar{\P}\, \P\,, \\
      &= m \, \int\, d{\tilde{k}} \left( a^{s\, \dag}_{\Vec{k}}\, a^{s}_{\Vec{k}}\,  + \,  b^{s\, \dag}_{\Vec{k}}\, b^{s}_{\Vec{k}}\right)\,,
\end{aligned}
\end{equation}
where $a^{s}_{\Vec{k}}$ and $b^{s}_{\Vec{k}}$ annihilate the lowest energy vacuum state $\ket{0}$\,, hence $H\, \ket{0} = 0$\,. Therefore, all the excitations have positive energy. The momentum operator can be computed easily as
\begin{equation}
\begin{aligned}
P^i &=i\int d^3 x\, \P^\dagger\,\p^i\P\,,\\
&=\int d\tilde{k}\left( a^{s\, \dag}_{\Vec{k}}\, a^{s}_{\Vec{k}}\,  + \,  b^{s\, \dag}_{\Vec{k}}\, b^{s}_{\Vec{k}}\right)\, k^i\,.
\end{aligned}
\end{equation}
and also, we would like to examine the charge that corresponds to the Carrollian boost:
\begin{equation}
\begin{aligned}
C^i&=\int d^3 x\, x^i m\bar{\P}\P\,, \\
&=-i\,m\(2\pi\)^3 \int d\tilde{k}d\tilde{l}\,\left(\ a^{s\,\dag}_{\vec{k}}a^{s}_{\vec{l}}+b^s_{\vec{k}} b^{s\,\dag}_{\vec{l}} \right) \partial^i\d^3(\vec{k}-\vec{l})\,. 
\end{aligned}
\end{equation}
where $\partial_i$ is derivative with respect to $k^i$\,. One can also obtain the  commutation relation as 
\begin{equation}
\begin{aligned}
[P_i, C_j] &= i\,\delta_{ij}H\,,
\end{aligned}
\end{equation}
where we used the definition of the generators and anti-commutation relation \eqref{19}\,. Finally, after discussing spacetime symmetries, we turn to the internal symmetries of the Carrollian Dirac field. Given that the Carrollian Dirac Lagrangian exhibits a global $U(1)$ symmetry,  it possesses a corresponding conserved charge, given by
\begin{equation}
\begin{aligned}
Q&=\int d^3 x\,\P^{\dag}\P\,,\\
&=\int d\tilde{k}\(a^{s\, \dag}_{\Vec{k}}\, a^{s}_{\Vec{k}}\,  - \,  b^{s\, \dag}_{\Vec{k}}\, b^{s}_{\Vec{k}}\)\,. 
\end{aligned}
\end{equation}
It is nothing but the electric charge. In addition to the global charge, the Lagrangian also exhibits a local\,, spatially dependent symmetry given by
\bea
\psi \rightarrow \psi' = e^{i\,\alpha(x)}\, \psi\,,
\eea 
where $\alpha(x)$ is a function of spatial coordinates only. This type of symmetry is related to supertranslations and is also a symmetry of complex Carrollian scalar theory {\cite{Cotler:2024xhb}}\,.  Since the Hamiltonian is bounded from below like the Carrollian scalar field theory, we can now define one particle states as
\begin{equation}
\begin{aligned}
	\ket{k\,, s} = \sqrt{2 m}\,a^{s\dag }_{\Vec{k}} \ket{0}\,,
\end{aligned}
\end{equation}
and as a result, we can define an inner product
\begin{equation}
\begin{aligned}
	\bra{ l\,, s'} \ket{k\,, s}= 2 m (2\pi)^3\, \delta^{(3)}(\vec{l}- \vec{k})\, \delta^{s\,,s'}\,,
\end{aligned}
\end{equation}
which is a Carrollian invariant object.  Now that we have all requisite information pertaining to the Carrollian Dirac field, we are now positioned to calculate the 2-point functions as follows
\begin{equation}
\begin{aligned}
		\bra{0}\P(t,\Vec{x})\Bar{\P}(t^{\prime},\Vec{y})\ket{0}=&\frac{e^{-i m\(t-t^{\prime}\)}}{2}\(\g^0+\mathbb{1}\) \delta^d(\Vec{x}-\Vec{y})\,,
\end{aligned}
\end{equation}  
and similarly, one can also show that 
\begin{equation}
\bra{0}\bar{\P}(t^{\prime},\Vec{y})\P(t,\Vec{x})\ket{0}=\frac{e^{i m\(t-t^{\prime}\)}}{2}\(\g^0-\mathbb{1} \)\delta^d(\Vec{x}-\Vec{y})\,. 
\end{equation}  
Consequently, the Feynman propagator can be obtained by the same procedure as 
\begin{equation}
\begin{aligned}
		S_F(t-t^{\prime},\Vec{x}-\Vec{y}) = &  	\bra{0}T\, \P(t,\Vec{x})\Bar{\P}(t^{\prime},\Vec{y})\ket{0}\, \\
		 =&\th(t-t^{\prime})\bra{0}\P(t,\Vec{x})\Bar{\P}(t^{\prime},\Vec{y})\ket{0}-\th(t^{\prime}-t) \bra{0}\bar{\P}(t^{\prime},\Vec{y})\P(t,\Vec{x})\ket{0}\,, \\
		=&\delta^d(\Vec{x}-\Vec{y})\[\th(t-t^{\prime})e^{-i m(t-t^{\prime})}\frac{\g^0+\mathbb{1}}{2}-\th(t^{\prime}-t)e^{i m(t-t^{\prime})}\frac{\g^0-\mathbb{1}}{2}\]\,, 
\end{aligned}\label{FeynmanDirac}
\end{equation}
where $T$ is the time ordering operator, and also in the momentum basis, we have 
\begin{equation}
		S_{DF}(w)=i\frac{w\g^0+m\mathbb{1}}{w^2-m^2+i\e}\,.
\end{equation}
In this section, we have achieved the canonical quantization of the Carrollian Dirac field. We are now ready to discuss the CPT transformations of the theory. 

%%%%%%%%%%%%%%%%%%%%%%%%%%%%%%%%%%%%%%%%%%%%%%%%%%
\subsection{CPT transformations} 
%%%%%%%%%%%%%%%%%%%%%%%%%%%%%%%%%%%%%%%%%%%%%%%%%%
In this section, we will explore the discrete symmetries of the Carrollian Dirac theory. CPT transformations are important in quantum field theory because they embody fundamental symmetries of nature, ensure the consistency of the theory, and provide a deep connection between the properties of particles and antiparticles. Given that any physical theory must be CPT  invariant, in the following, we commence by formulating the charge conjugation, parity, and time reversal transformations related to Carrollian Dirac theory, respectively \footnote{Note that CPT transformations in two and three dimensions are discussed in \cite{Banerjee:2022ocj, Hao:2022xhq} in different context.}.
%%%%%%%%%%%%%%%%%%%%%%%%%%%%%%%%%%%%%%%%%%%%%%%%%%
\subsubsection{Charge conjugation}
%%%%%%%%%%%%%%%%%%%%%%%%%%%%%%%%%%%%%%%%%%%%%%%%%%
We begin our discussion with the charge conjugation transformation. Charge conjugation defines the symmetry between particles and antiparticles and is implemented by a unitary operator $C$.  This operator transforms a fermion into its corresponding antifermion while leaving its spin unchanged. Therefore, we can deduce the following transformations :
\begin{equation}
    C\, a^s_p C= b_p^s\,, \qquad   C\, b^s_p C= a_p^s\,. 
\end{equation}
Then, in this context, one of our fundamental requirements is the flipped spinor, which is defined as follows :
\begin{equation}
    \xi^{-s}=-i \s^2\(\xi^s\)^*\,, \label{25}
\end{equation}
where $\s^i$ are Pauli matrices. Noting that since  $\eta^s=\xi^{-s}$, then we have 
\begin{equation}
(v^s)^*=\begin{pmatrix} \sqrt{m}\(-i\s^2   \(\xi^s\)^*\) \\
-\sqrt{m}\(-i\s^2 \(\xi^s\)^*\)
\end{pmatrix}^*=\sqrt{m}\begin{pmatrix} -i\s^2\xi^{s}\\
i\s^2\xi^{s}
\end{pmatrix}=\begin{pmatrix}
    -i\s^2& 0\\
    0& i\s^2
\end{pmatrix}u^s\,,\\
\end{equation}  
and we may also articulate it in an alternate form:  
\begin{equation}
(v^s)^*=\begin{pmatrix}
0& -i\s^2\\
i\s^2& 0
\end{pmatrix}u^s\,.   \label{26}
\end{equation}  
Therefore, we can deduce the following relations between  $u^s$ and $v^s$ as 
\begin{equation}
\(u^s\)^*= \sqrt{m}\begin{pmatrix}
i\s^2(\eta^s)^*\\
i\s^2(\eta^s)^*
\end{pmatrix}^*=\begin{pmatrix}
i\s^2&0\\
0&-i\s^2
\end{pmatrix}v^s\,.
\end{equation}  
But also, we have 
\begin{equation}
(u^s)^*=\begin{pmatrix}
    0&-i\s^2\\
    i\s^2&0
\end{pmatrix}v^s\,.  \label{27}
\end{equation} 
Then, we can continue with the field  $\P$ as
\begin{equation}
\begin{aligned}
C\P C=&\int d\Tilde{k}\(e^{-i m t} e^{i\Vec{k}\cdot\Vec{x}}C a^s_{\Vec{k}} C u^s+ e^{i m t} e^{-i\Vec{k}\cdot\Vec{x}}C b^{s\dag}_{\Vec{k}}C v^s \)\,,\\
=&\begin{pmatrix}
    0&-i\s^2\\
    i\s^2&0
\end{pmatrix} \P^*=-i\g^2\P^*\,.
\end{aligned}\label{28}
\end{equation} 
Similarly, $C\bar{\P}C$=$i \P^T \g^2\g^0$. Given the transformation of these two elementary quantities, we are permitted to compute the bilinears that may potentially be incorporated into the Lagrangian:
\begin{equation}
\begin{aligned}
C\Bar{\P}\P C=& C\Bar{\P}C C\P C=\P^T\g^2 \g^0\g^2\P^*=\P^T\g^0\P^*= \Bar{\P}\P\,,\\
C i\Bar{\P}\g^5\P C=&i\(i\P^T\g^2\g^0 \) \g^5 \(-i\g^2\P^*\)=-i\P^T\g^0\g^5\P^*=i\Bar{\P}\g^5\P\,, \\
C i\Bar{\P}\g^0\Dot{\P}C=&i C\Bar{\P} C\g^0 C \Dot{\P}C=i\(i\P^T\g^2\g^0 \)\g^0\(-i\g^2\Dot{\P}^*\)\,,\\
=&-i\P^T\Dot{\P}^*=-i\Dot{\P}^\dag\P=-i\partial_t\(\Bar{\P}\g^0\P\)+i\Bar{\P}\g^0\Dot{\P}\,.  \label{29}
\end{aligned}
\end{equation}  
Given that boundary terms vanish, we have the same term. Thus, the terms within our action exhibit invariance under charge conjugation. Another bilinear that might be worth to examine is $\Psi^{\dag}\Psi$ which resulted as
\begin{equation}
\begin{aligned}
C\P^{\dag}\P C=&C\P^{\dag}C C\P C=i\P^T\g^2\(-i\g^2\P^*\)=-\P^T\P^*\,,\\
=&-\P^{\dag}\P\,, 
\end{aligned} \label{30}
\end{equation}
which is not invariant under charge conjugation, albeit invariant under Carroll transformations.

%%%%%%%%%%%%%%%%%%%%%%%%%%%%%%%%%%%%%%%%%%%%%%%%%%
\subsubsection{Parity}
%%%%%%%%%%%%%%%%%%%%%%%%%%%%%%%%%%%%%%%%%%%%%%%%%%
Now, we proceed by investigating the parity transformation of the Carrollian Dirac field. The parity operator transforms the state $a^{s\dag}_{\Vec{k}} \ket{0}$ into $a^{s\dag}_{-\Vec{k}} \ket{0}$ as we know from the Lorentzian case. Consequently, we may interpret the operator as follows: 
\begin{equation}
P a^{s}_{\Vec{k}}P=\eta_a a^s_{-\Vec{k}}\hspace{0.1cm}\And \hspace{0.1cm} P b^{s}_{\Vec{k}}P=\eta_b b^s_{-\Vec{k}}\,,
\end{equation}
where $\eta_a$ and $\eta_b$ represents possible phase factors. Thus,
\begin{equation}
\begin{aligned}
P \P(t,\Vec{x}) P&=\int \frac{d\Tilde{k}}{\sqrt{2m}} \(e^{-i m t}e^{i\Vec{k}\cdot\Vec{x}} \eta_a a^s_{-\Vec{k}} u^s+ e^{i m t}e^{-i\Vec{k}\cdot\Vec{x}}\eta_b^* b^{s\dag}_{-\Vec{k}} v^s \)\,.
\end{aligned}\label{32}
\end{equation}
After renaming the spatial momenta  $\Vec{k}$ as $-\Vec{k^{\prime}}$, \eqref{32} turns out to be 
\begin{equation}
\begin{aligned}
P \P(t,\Vec{x}) P&=\int \frac{d\Tilde{k^{\prime}}}{\sqrt{2m}} \[e^{-i m t}e^{-i\Vec{k^{\prime}}\cdot\Vec{x}}\eta_a a^s_{\Vec{k^{\prime}}}u^s+e^{i m t}e^{i\Vec{k^{\prime}}\cdot\Vec{x}}\eta_b^* b_{\Vec{k^{\prime}}}^{s\dag}v^s \]\,. 
\end{aligned}\label{33}
\end{equation}
Interestingly, in the Carrollian Dirac field, the parity operator introduces only a phase factor, unlike in the relativistic Dirac field. Specifically, if $\eta_a=\eta_b^*$,  we find that $\eta_a \eta_b=1=\eta_b\eta_b^*$. In contrast, in the relativistic case, the product of these phase factors equals $-1$\,. This difference arises because, in the Carrollian theory, the spinor basis is independent of the spatial momenta (see Eq. \eqref{22}). This highlights a fundamental distinction between the relativistic Dirac field and the Carrollian Dirac field\,.  As a result \eqref{33} becomes
\begin{equation}
P\P(t,\Vec{x})P=\eta_a \P(t,-\Vec{x})\,.
\end{equation}
Similarly, $P\Bar{\P}(t,\Vec{x})P=\eta_a^*\Bar{\P}(t,-\Vec{x})$. We are now prepared to compute the transformations of bilinears that are of interest to us:
\begin{equation}
\begin{aligned}
P\Bar{\P}\P P=&P\Bar{\P}P P\P P= \abs{\eta_a}^2\Bar{\P}\P(t,-\Vec{x})=\Bar{\P}\P(t,-\Vec{x})\,,\\
P\P^{\dag}\P P=&\P^{\dag}\P(t,-\Vec{x})\,,\\
P i\Bar{\P}\g^5\P P=&i \Bar{\P}\g^5\P(t,-\Vec{x})\,,\\
P i\Bar{\P}\g^0\Dot{\P}P=&i\Bar{\P} \g^0 \Dot{\P}(t,-\Vec{x})\,.
\end{aligned}\label{35}
\end{equation}

%%%%%%%%%%%%%%%%%%%%%%%%%%%%%%%%%%%%%%%%%%%%%%%%%%
\subsubsection{Time reversal}
%%%%%%%%%%%%%%%%%%%%%%%%%%%%%%%%%%%%%%%%%%%%%%%%%%
As a final exploration of discrete symmetry transformations, we shall investigate the time reversal transformation and its consequences. Essentially,  electric Carrollian field theory can be considered as a one-dimensional quantum mechanics \cite{Kasikci:2023zdn}. Consequently, the time reversal operator remains an antiunitary operator in Carrollian field theory as well. To facilitate this inquiry, we will use flipped spinors \eqref{25} as demonstrated below \cite{Peskin:1995ev}:
\begin{equation}
u^{-s}=\sqrt{m}\begin{pmatrix}
    \xi^{-s}\\
    \xi^{-s}
\end{pmatrix}=\sqrt{m}\begin{pmatrix}
    -i\s^2\(\xi^s\)^*\\
    -i\s^2\(\xi^s\)^*
\end{pmatrix}=-i\begin{pmatrix}
    \s^2&0\\
    0&\s^2
\end{pmatrix}\(u^s\)^*=-\g^1\g^3\(u^s\)^*\,. \label{36}
\end{equation}
Similarly, $v^{-s}=-\g^1\g^3\,(v^s\,)^{*}$\,. As a next step, we define the time reversal transformation of fermion annihilation and anti-fermion creation operators as
\begin{equation}
    T a^{s}_{\Vec{k}} T=a^{-s}_{-\Vec{k}}\And T b^{s\dag}_{\Vec{k}} T=b^{-s\dag}_{-\Vec{k}}\,.
\end{equation}
Note that time reversal reverses the  momentum and flips the  spin.  Consequently, we are allowed to  obtain the time reversal transformation of $\P$:
\begin{equation}
\begin{aligned}
T\P(t,\Vec{x})T&=\int \frac{d\Tilde{k}}{\sqrt{2m}} \(e^{i m t}e^{-i \Vec{k}\cdot\Vec{x}}a^{-s}_{-\Vec{k}}\(u^s\)^*+e^{-i m t}e^{i\Vec{k}\cdot\Vec{x}}b^{-s\dag}_{-\Vec{k}}\(v^s\)^*  \)\,,\\
&=-\g^3\g^1 \P(-t,\Vec{x})\,. 
\end{aligned}
\end{equation}  
Likewise, $T\Bar{\P}(t,\Vec{x})T$=$-\P^{\dag}(-t,\Vec{x}) \g^1\g^3 \g^0$=$-\Bar{\P}(-t,\Vec{x})\g^1 \g^3 $. In the ensuing step, we compute the bilinears that we are interested in charge conjugation and parity transformations:
\begin{equation}
\begin{aligned}
T\Bar{\P}\P(t,\Vec{x})T=&T\Bar{\P}(t,\Vec{x})T T\P(t,\Vec{x})T=-\Bar{\P}\g^1\g^3\(-\g^3\g^1\P\)=\Bar{\P}\P(-t,\Vec{x})\,, \\
T\P^{\dag}\P(t,\Vec{x})T=&\P^{\dag}\P(-t,\Vec{x})\,, \\
Ti\Bar{\P}\g^5\P(t,\Vec{x})T=&-i\(-\Bar{\P}\g^1\g^3\)\g^5\(-\g^3\g^1\P\)=-i\Bar{\P}\g^5\P(-t,\Vec{x})\,, \\
T i\Bar{\P}\g^0\Dot{\P}(t,\Vec{x})T=&-i\(-\Bar{\P}\g^1\g^3\g^0\)\(\g^3\g^1\Dot{\P}\)=i\Bar{\P}\g^0\Dot{\P}(-t,\Vec{x})\,.
\end{aligned}
\end{equation}

\begin{table}[ht]
\centering
\scalebox{1.2}{
\begin{tabular}{|c|c|c|c|c|}
\hline
 & $\Bar{\P}\P$ & $\P^{\dag}\P$ & $i\Bar{\P}\g^5\P$ & $i\Bar{\P}\g^0\Dot{\P}$ \\
\hline
$C$ & +1  & -1  & +1 &+1 \\
\hline
$P$ & +1  &+1  &+1 &+1\\
\hline
$T$ & +1  &+1  &-1 &+1\\
\hline
$C P T$ &+1  &-1  &-1 &+1\\
\hline
\end{tabular}
}
\caption{CPT transformations of the bilinears}
\label{table1}
\end{table}
We can succinctly encapsulate our findings regarding CPT  transformations as given in Table \ref{table1}\,. As can be seen from the table,  the only permitted terms in the Lagrangian are  $i\Bar{\P}\g^0\Dot{\P}$ and the mass term  $\bar{\Psi}\Psi$. 
%%%%%%%%%%%%%%%%%%%%%%%%%%%%%%%%%%%%%%%%%%%%%%%%%%
\section{Carrollian Yukawa theory}
%%%%%%%%%%%%%%%%%%%%%%%%%%%%%%%%%%%%%%%%%%%%%%%%%%
In this section, we will consider the Yukawa interaction as a toy model to unreveal the structure of the electric Carrollian Dirac fermion and its interaction with the scalar field.  We will use the perturbative approach when dealing with the interaction. We first seek the tree-level result and then we are going to discuss the one-loop contributions of the vertices and self-energy diagrams. Finally, we calculate the beta functions for Carrollian Yukawa theory and investigate the non-trivial fixed points and stability analysis.

The electric Carrollian Yukawa theory can be obtained from its relativistic counterpart by applying similar scalings to the Yukawa interaction term, given by $\tilde{g} \tilde{\phi}\bar{\tilde{\Psi}}\tilde{\Psi} = g\phi\bar{\Psi}\Psi$ with $\tilde{g} = g/c^2$\,. This results in the Carrollian Yukawa Lagrangian\footnote{The first classical Carrollian Yukawa theory with massless fermions and a massless scalar field is examined in \cite{Bagchi:2019xfx}\,. }, which includes the Carrollian scalar field, a scalar self-interaction term $\lambda \phi^4$, the Carrollian Dirac field, and their interaction term, given by
\begin{equation}
\mathcal{L}_{CY}= \Bar{\P}i\g^0\dot{\P}-m\Bar{\P}\P+\frac{1}{2}\dot{\f}^2-\frac{M^2}{2}\f^2- g\f\Bar{\P}\P -\frac{\l}{4!}\f^4\,.\label{45}
\end{equation}
Despite the primary motivation of this investigation being the derivation of the $\b$ function, it is of interest to explore the tree-level Yukawa potential term associated with this theoretical framework. Therefore, in addition to the scalar Feynman rules provided in  \eqref{scalarfeynmanrules}\,, let us explicitly present the tree-level Feynman Rules for Dirac field and Yukawa interaction in the momentum basis as follows:
\begin{equation} 
\begin{aligned}
        \text{Dirac propagator:}\qquad &  S_{DF}(\slashed{w})&=&\frac{i}{\slashed{w}-m+i\e}\,,  \\
        	\text{Yukawa vertex:}\qquad & V_g  &=&  -i g\,,
\end{aligned}
\end{equation}
where $\slashed{w}=w\g^0$\,, $S_{DF}$ denotes the Dirac propagator\,, and $V_g$ denotes the fermion-scalar vertex\, respectively\,. From now on, the solid line will represent the Dirac field, while the dashed line will indicate the scalar field\,. 
\begin{figure}[ht]
    \centering
\includegraphics[scale=.2]{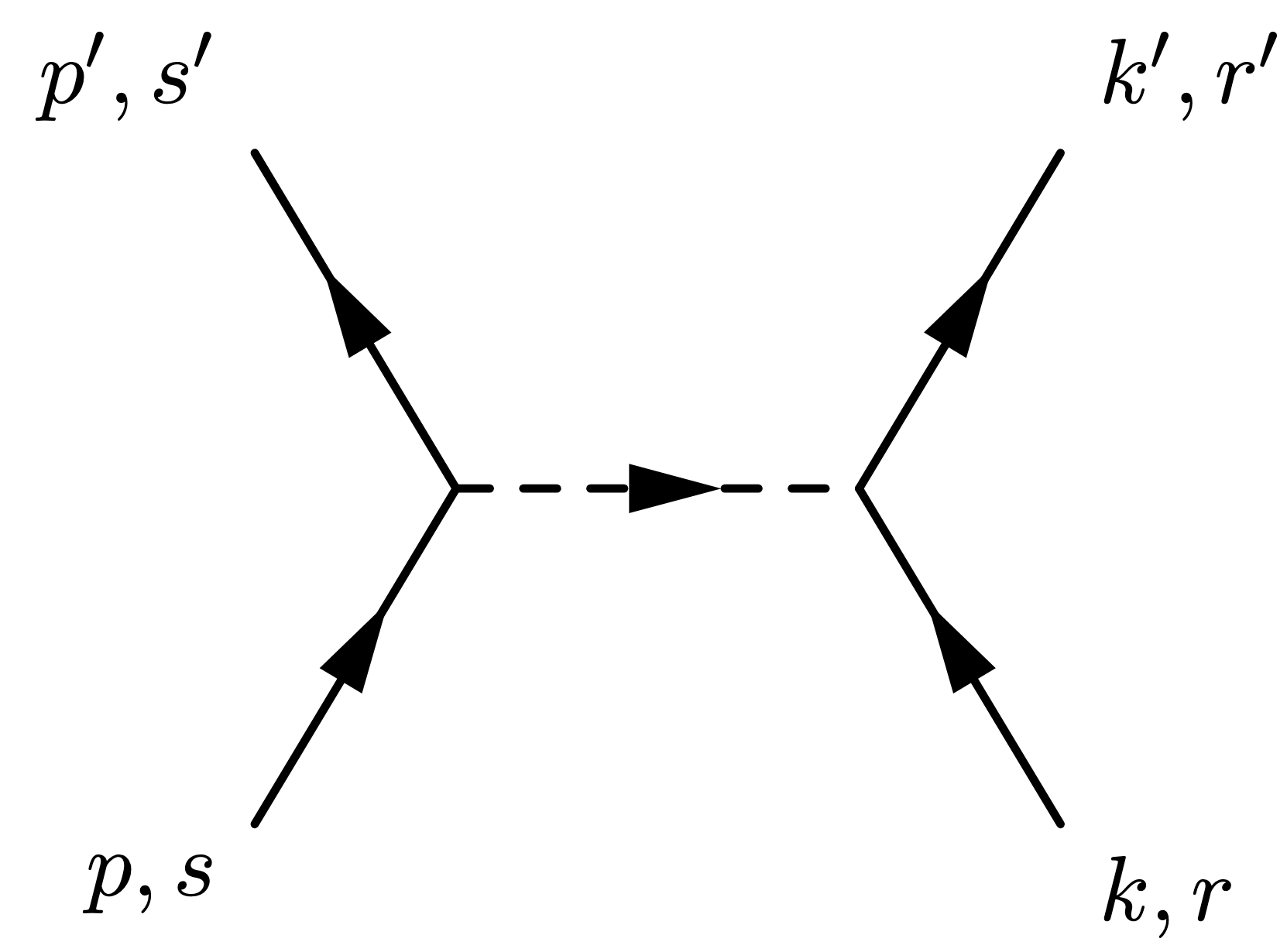}
\caption{Yukawa vertex\,.}\label{yukawatree}
\end{figure}
In order to elucidate the Yukawa potential, we focus on the scattering amplitude of the tree-level Feynman diagram in Figure \ref{yukawatree} via adopting the notation of \cite{Peskin:1995ev,Schwartz:2014sze}\,. 
Since each vertex corresponds to $-ig$ in the momentum basis, then the Feynmann diagram corresponds to
\begin{equation}
\begin{aligned}
i\mathcal{M}=&(-i g)^2\Bar{u}^{s^{\prime}}(p^{\prime}) u^s(p) \frac{i}{\Tilde{w}^2-M^2+i\e}\Bar{u}^{r^{\prime}}(k^{\prime}) u^r(k)\,,\\
=&\frac{-i g^2}{\Tilde{w}^2-M^2}(2m\delta^{s s^{\prime}})(2m\delta^{r r^{\prime}})\,.
\end{aligned}
\end{equation}
%Since $\Tilde{w}$ = $m_{\Bar{u}}-m_u=0$. 
where $\Tilde{w}=w_p-w_{p^{\prime}}$. Comparing with the Born approximation, we see that Yukawa potential in momentum basis can be written as
\begin{equation}
\Tilde{V}(w,\Vec{p})=\frac{g^2}{w^2-M^2}\,. \hspace{0.5cm}
\end{equation}
Inverting the Fourier transform to find $V(t,\vec{x})$:
\begin{equation}
\begin{aligned}
V(t,\Vec{x})=&\delta^3(\Vec{x})\int \frac{d w}{2\pi} \frac{g^2 e^{-i w t}}{\(w-M+i\e \) \(w+M-i\e \)}\,,\\
=&\delta^3(\Vec{x})\frac{g^2}{2\pi}\[\th(t)\frac{e^{-i Mt}}{2 M}\(-2\pi i\)+\th(-t)\frac{e^{i Mt}}{-2 M}\(2\pi i\) \]\,,\\
=&-i\frac{g^2}{2 M}e^{-i M\abs{t}}\delta^3(\Vec{x})\,,
\end{aligned}
\end{equation}
which means we indeed have an \textit{ultralocal interaction}. This type of interaction is known as a Dirac-delta potential with time-dependent factor and has been widely studied in quantum physics, quantum tunnelling, quantum chaos, quantum defects, spectroscopy, and quantum optics (see \cite{martinez2001transmission, Erman2020:xyz, Mudra:2021nez} for details and references therein). It is known that electric Carrollian field theories consist of an infinite collection of one-dimensional quantum mechanical systems \cite{Kasikci:2023zdn}. This analysis, therefore, provides a Carrollian origin for point-like interactions in quantum physics.
%%%%%%%%%%%%%%%%%%%%%%%%%%%%%%%%%%%%%%%%%%%%%%%%%%
\subsection{Wilsonian Renormalization Group}
%%%%%%%%%%%%%%%%%%%%%%%%%%%%%%%%%%%%%%%%%%%%%%%%%%
In this section, we use the Wilsonian renormalization group (RG) framework to study Carrollian Yukawa theory. The Wilsonian RG flow describes how the couplings of the theory evolve with the energy scale. While our focus in this paper is on the flow of the couplings, one could also apply dimensional regularization and other conventional renormalization techniques in quantum field theory by introducing counterterms. However, due to the distinct scalings of time and spatial coordinates, dimensional regularization must be handled carefully. Moreover, the renormalization conditions for Carrollian field theory are still not well established and require further study. Therefore, we leave these issues for future work and instead follow \cite{Banerjee:2023jpi}, which applied the RG flow approach to Carrollian scalar field theory. 
\subsubsection{Warming up: Relativistic theory}

The Wilsonian renormalization is a powerful conceptual and technical framework in theoretical physics, particularly in quantum field theory and statistical mechanics, for understanding how physical theories behave at different energy scales. In Wilsonian renormalization, the RG flow describes how parameters of the theory scale as the high energy degrees of freedom are integrated out. The key idea is to systematically integrate out these high energy modes to derive an effective field theory that describes the physics at lower energy scales\, \cite{Wilson:1971bg, Wilson:1971dh}.

In the path integral formulation of quantum field theory (QFT), Wick rotated  partition function is expressed as an integral over all possible field configurations
\begin{equation}
Z = \int\, D \Phi\, e^{-\, S[\Phi]}\,, \label{pathintegral}
\end{equation}
where $\Phi$ represents the all the fields (scalar field $\phi$ and Dirac field $\psi$) and the $S[\Phi]$ is the corresponding action. To integrate the high energy modes, we split the field $\Phi$ into high energy $\Phi_H$ and low energy modes $\Phi_L$ as
\bea
\Phi= \Phi_H + \Phi_L\,.
\eea
The systematic is simply introduce the upper cut-off $\L$ such that $\Phi=0$ if $\abs{k}>\L$ and with an intermediate cut-off\, $\frac{\L}{b}$ where $b>1$\,. Then
\begin{equation}
\begin{aligned}
\Phi&=\Phi_L \quad\text{if}\quad |k|<\, \frac{\L}{b},\quad \text{else}\quad \Phi_H\,,
\end{aligned}
\end{equation}
Thus, the path integral \eqref{pathintegral} becomes
\begin{equation}
Z = \int\, D \Phi_H\, D\Phi_L\, e^{-\, S[\Phi_H + \Phi_L]}\,,
\end{equation}
If we perform the integral over high energy modes $\Phi_H$, then 
\begin{equation}
Z = \int\, D\Phi_L\, e^{-\, S_{eff}[\Phi_L]}\,,
\end{equation}
where the effective action $S_{eff}[\Phi_L]$ is obtained  by integrating out $\Phi_H$\,. After integrating out the high-energy modes, the effective action  $S_{eff}[\Phi_L]$ typically contains new terms that were not present in the original action. These terms represent the influence of the high-energy modes on the low-energy physics. To compare the effective theory at different scales, we rescale the momenta and fields so that the cutoff $\Lambda$ is restored to its original value. This rescaling step is crucial for understanding how the parameters of the theory (e.g., couplings, masses) change with scale, a process known as renormalization group flow.

First, we extract the scalings of fields and mass parameters from the free action. The free action defines the system at a reference scale (e.g., the cutoff $\Lambda$) and consists of only the quadratic (Gaussian) part, with the bare mass ( $M_0$ and $m_0$ for Yukawa model)\,. It describes the non-interacting (Gaussian) fixed point, which is a natural starting point for RG analysis since it approximates high-energy behavior. Interactions can then be treated as perturbations. The free action also $S_0$\, determines the canonical scaling dimensions of fields and parameters, which are essential for understanding RG flow. These dimensions follow from requiring $S_0$ to remain invariant under rescaling, establishing a baseline before including interactions.

To extract the scaling properties of the fields and mass parameters at leading order, we begin with the Wick rotated free relativistic action\footnote{In this section, we will not use  $\tilde{X}$ for relativistic fields for convenience as it is understood from the context.}:
\begin{equation}
S_0 =\int d^4 x \[ \frac{\f}{2}\(-\p^2 + M_0^2\)\f+\bar{\P}\(\slashed{\p} + m_0\)\P \]\,, 
\end{equation}
where $M_{0}$ and $m_0$ indicate that these parameters are defined at the initial energy scale of the theory. To study the scaling behavior, we rescale the intermediate energy scale $k$ and coordinates $x$ as follows
\bea
k'^{\mu}=b\, k^\mu \And x'^{\mu}=x^\mu/b\,,
\eea
where $b$ is the scaling factor and $k^\mu$ and $x^\mu$ are four-momentum and four-dimensional spacetime coordinates, respectively. Accordingly, free action becomes 
\begin{equation}
S_0= \int d^4x' \[\frac{b^2\f}{2}\(-\p'^2\, + \, b^2 M_{0}^2\)\f+b^3\bar{\P} \(\slashed{\p}'\, + b\, m_0\, \)\,\P\]\,.
\end{equation}
We see that the kinetic terms remain canonical if $\f'$ = $b \f$ and $\P'$=$b^{3/2} \P$\,. This gives the engineering (canonical) dimension of the fields, namely, the scaling dimension of $\f$ ($d_\phi$) is equal to 1 and the scaling dimension of $\P$ ($d_\Psi$) is equal to 3/2. Additionally, we observe that the scalar mass scales as $b^2\, M_{0}$ and the fermion mass scales as $ b\,m_{0}$\,.

\begin{figure}[ht]
    \centering
\includegraphics[scale=.23]{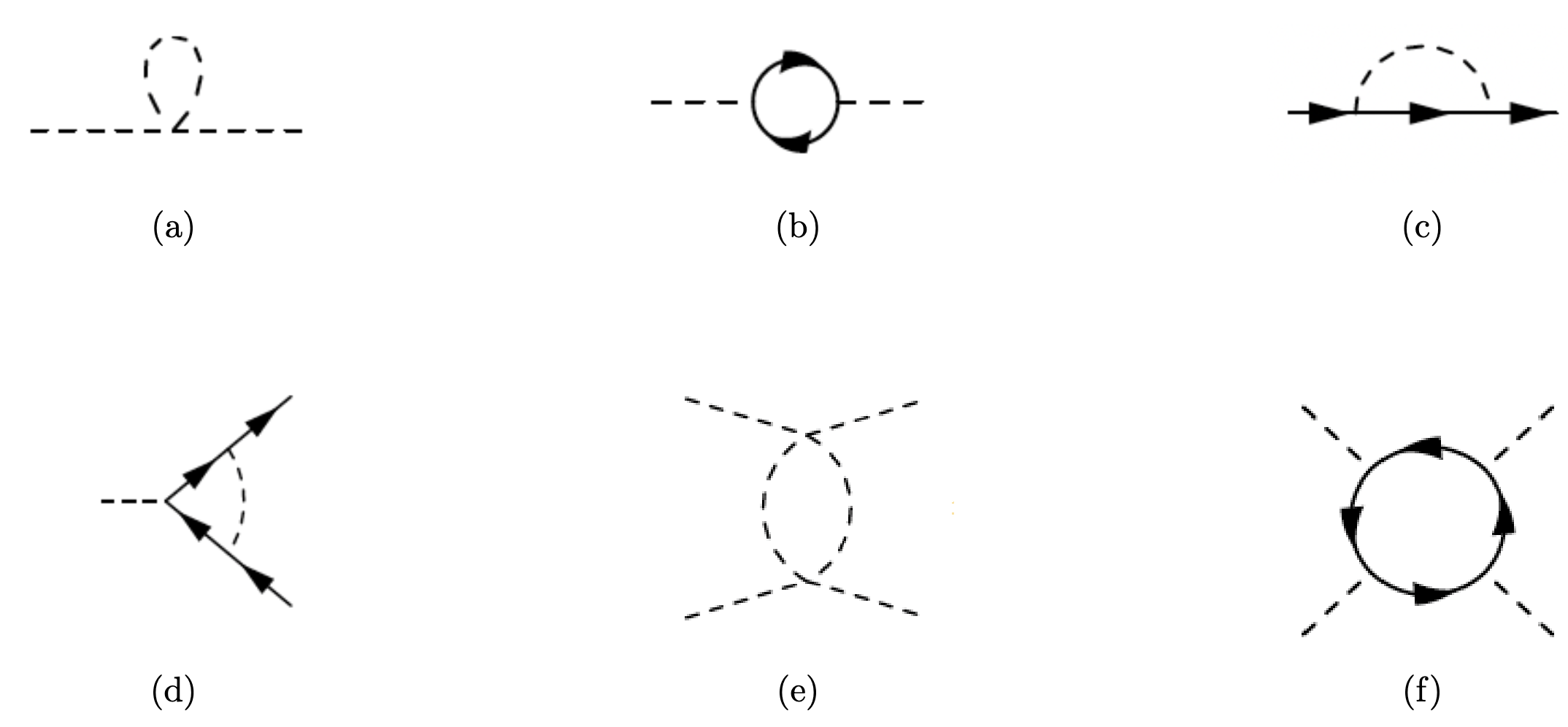}
\caption{Here diagrams (a) and (b) illustrate the scalar's self-energy diagrams resulting from the couplings $\l$ and $g$, respectively. Diagram (c) presents the self-energy diagram for the fermion attributed to the coupling $g$. Diagram (d) depicts the one-loop Yukawa vertex, whereas diagrams (e) and (f) correspond to the one-loop $\l$ vertex arising from the couplings $\l$ and $g$, respectively.}\label{fig:loop}
\end{figure}

When interactions are introduced\,, the RG flow of parameters can be computed perturbatively around the Gaussian fixed point. In order to understand the scaling of the couplings, we have to find the scaling dimensions of the couplings. Since it is well-known that any operator $\mathcal{O}$ scales as $ b^{4-d_{\mathcal{O}}}$ for 1+3 dimensional systems, we obtain:
\begin{equation}
\begin{aligned}
\frac{\l}{4!}\f^4&\rightarrow&\frac{\l}{4!}\f^4 b^{4-4d_\f}&=&\frac{\l}{4!}\f^4\,,\\
g\f\bar{\P}\P&\rightarrow &g\f\bar{\P}\P b^{4-d_\f-2d_\P}&=&g\f\bar{\P}\P\,. 
\end{aligned}
\end{equation}
Thus, the interacting action \eqref{45} in Euclidean time reads \footnote{See \cite{Peskin:1995ev} for a discussion of  these terms in the scalar $\lambda\,  \phi^4$ theory.}:
\begin{equation}
\mathcal{L}_{CY}=  \Bar{\P}\g^\mu\,\partial_\mu \P + m(b)\, \Bar{\P}\P+\frac{1}{2}\dot{\f}^2 + \frac{M^2(b)}{2}\f^2 + g(b)\f\Bar{\P}\P + \frac{\l(b)}{4!}\f^4\,, \label{EuclideanYukawa}
\end{equation}
where 
\begin{equation}
\begin{aligned}
M^2(b)=&\frac{b^2\[M_{0}^2+\d_{M}\]}{1 + \delta_\phi}\,,\quad m(b)=\frac{b\[m_0+\d_m\]}{1 + \delta_\psi }\,, \\
\l(b)=& \frac{\l_0+\d_\l}{(1 + \delta_\phi)^2}\,\hspace{1.45cm} g(b)= \frac{g_0+\d_g}{(1 + \delta_\psi)(1 + \delta_\phi)^{1/2}}\,.
\end{aligned}\label{parametersRel}
\end{equation}
Note that $\delta$ terms given in the expression \eqref{parametersRel} account for the quantum contribution arising from the loop diagrams. In the sequel, we will follow the conventional path to connect these $\delta$ terms at one-loop order with the corresponding Feynman diagrams given by the Figure \eqref{fig:loop}.  To begin our analysis, we compute the field strength $(\delta_\phi\,, \delta_\psi)$\,, and mass renormalization $(\delta_M\,, \delta_m)$\,, arising from the self-interaction and Yukawa interaction, respectively. For convenience, first we adopt relativistic notation. Thus, the first term is given by the expressions corresponding to diagrams (a) and (b) in Figure \ref{fig:loop}: 
\begin{equation}
\begin{aligned}
&i\,\left( p^2\,\delta_\phi -\d_M\right)= -\frac{i\l}{2}\int \frac{d^4 k}{\(2\pi\)^4}\frac{i}{k^2-M^2}  -(-i g)^2\int \frac{d^4 k}{\(2\pi\)^4} \text{Tr}\[\frac{i}{( \slashed{k} - \slashed{p}-m)}\, \frac{i}{( \slashed{k} -m)}  \]\,. \label{loop1rel}
\end{aligned}
\end{equation}
This term corresponds to field strength and mass renormalization for the scalar field. The field strength and mass renormalization term for Dirac field is given by the diagram (c) in Figure \ref{fig:loop}:
\begin{equation}
\begin{aligned}
i\left(\slashed{p}\, \delta_\psi - \delta_m\right) = (- i g)^2\, \int \frac{d^4 k}{\(2\pi\)^4}\frac{i}{\slashed{k}-m}\, \frac{i}{(p-k)^2 - M^2}\,. \label{loop1rel2}
\end{aligned}
\end{equation}

The vertex diagrams also have contributions at one-loop level as seen by the Figure- \ref{fig:loop}. Firstly, let us consider the contribution to the Yukawa vertex  (given by diagram (d)). The corresponding term is then given by
\begin{equation}
\begin{aligned}
&-i\d_g=(-i g)^3\int \frac{d^4 k}{\(2\pi\)^4}\frac{i^3}{k^2-M^2}\frac{k^2+m^2}{\(k^2-m^2\)^2}\,.
\end{aligned}
\end{equation}
As a concluding contribution, we are particularly focused on $\d_\l$ term arising from the scalar self-interaction and fermionic loop given in diagrams (e) and (f) in Figure-\ref{fig:loop}. In the relativistic notation, we have the one-loop contribution the self interaction scalar vertex as the following form:  
\begin{equation}
\begin{aligned}
-i\d_{\l}=&\frac{3}{2}(-i\l)^2\int \frac{d^4 k}{\(2\pi\)^4}\frac{i^2}{\(k^2-M^2\)^2} -6 \(-ig\)^4\int \frac{d^4 k}{\(2\pi\)^4} \text{Tr}\[\frac{i^4(\slashed{k}+m)^4}{(k^2-m^2)^4}\]\,.\end{aligned}
\end{equation}  
From this point onward, we will consider the case of vanishing external momenta, $p=0$ for convenience. By setting the external momenta $p$ to zero, we exclude the contributions to field strength renormalization,  and we will therefore neglect such terms in \eqref{parametersRel} in the subsequent analysis. This is legitimate for several reasons: first, it significantly simplifies the analysis of the beta functions, and second, it aligns with the conventional approach to the beta function calculations, where one typically expands the expressions around $p=0$\,.  Then, the  field strength renormalization terms can be computed within this framework. A detailed discussion of the latter approach is provided in the Appendix\,.

Therefore, in terms of  Euclidean momentum $k^0_E=-i k^0$, the relevant terms for the beta function analysis are given by the following terms:

\begin{equation}
\begin{aligned}
\d_{M}&= \frac{\l}{16\pi^2}\int^{\L}_{\frac{\L}{b}}d k_E k_E^3\frac{1}{k_E^2+M^2} + \frac{g^2}{2\pi^2} \int^{\L}_{\frac{\L}{b}} d k_E k_E^3\frac{-k_E^2+m^2}{\(k_E^2+m^2\)^2}\,, \nn \\
\d_m&=-\frac{g^2 }{8\pi^2}\int^{\L}_{\frac{\L}{b}} d k_E k_E^3\frac{1}{k_E^2+M^2}\frac{m}{k_E^2+m^2}\,, \nn \\
\d_g&= \frac{g^3}{8\pi^2}\int^{\L}_{\frac{\L}{b}} d k_E k_E^3 \frac{-k_E^2+m^2}{\(k_E^2+M^2\)\(k_E^2+m^2\)^2}\,, \nn \\
\d_\l=& -\frac{3\l^2}{16\pi^2}\int^{\L}_{\frac{\L}{b}}  dk_E  k_E^3\frac{1}{\(k_E^2+M^2\)^2} + \frac{3 g^4 }{\pi^2}\int^{\L}_{\frac{\L}{b}}  dk_E  k_E^3 \frac{k_E^4+m^4-6m^2 k_E^2}{(k_E^2+m^2)^4}\,.
\end{aligned}
\end{equation}
Note that we explicitly show the integration boundaries for the $\beta$ function analysis. Up until now, we have calculated the one-loop contributions to the parameters of the theory, such as $M^2$\,, $m$\,, $\lambda$\,, and $g$, respectively\,. Having acquired all the requisite information, we may now proceed with the computation of the $\b$  function at the one-loop order. The $\b$ functions of the relativistic Yukawa theory can be obtained by noting that for any parameter $X_i = \left(M^2\,,  m\,, \lambda\,, g \right)$:
\begin{equation}
    \b_i=-b\frac{\p X_i}{\p b}|_{b=1}\,. 
\end{equation}
Using this definition and \eqref{parametersRel},  we obtain the $\b$ functions of the relativistic Yukawa theory as \footnote{For a detailed discussion of the Wilsonian approach to Yukawa theory with symmetry breaking term, see \cite{Clark:1992jr,Gies:2010mqh, Krajewski:2014vea}.}:

\begin{equation}
\begin{aligned}
\b_{M^2}&=-2M^2-\frac{\l}{16\pi^2}\frac{\L^4}{\L^2+M^2}+\frac{g^2 \L^4}{2\pi^2}\frac{\L^2-m^2}{\(\L^2+m^2\)^2}\,,\\
\b_m&=-m +\frac{g^2}{8\pi^2}\frac{\L^4}{\L^2+M^2} \frac{m}{\L^2+m^2}\,, \\
\b_{g}&=\frac{g^3}{8\pi^2}\frac{\L^4}{\L^2+M^2} \frac{\L^2-m^2}{\(\L^2+m^2\)^2}\,,\\
\b_{\l}&=\frac{3\l^2}{16\pi^2}\frac{\L^4}{\(\L^2+M^2\)^2}-\frac{3 g^4}{\pi^2}\frac{\L^4\(\L^4+m^4-6m^2\L^2\)}{\(\L^2+m^2\)^4}\,.
\end{aligned}\label{relbeta4}
\end{equation}  

In the Wilsonian RG framework, the Gaussian fixed point, nontrivial fixed points, and stability of fixed points play crucial roles in understanding phase transitions and universality classes of physical systems. A fixed point of the RG flow is a point in the space of couplings where the beta functions vanish:
\begin{equation}
    \b_i(g^*) = 0\,,
\end{equation}
where $g^*$ is a value of the coupling constant where the beta function vanishes. The Gaussian fixed point refers to the trivial or free theory where the interaction terms vanish. In the context of the RG flow, this corresponds to the point in parameter space where all couplings are zero (except for the kinetic term). The Gaussian fixed point is an important example, but there can be nontrivial fixed points where interactions remain finite. Nontrivial fixed points often correspond to critical points of phase transitions (e.g., Wilson-Fisher fixed point in $\lambda\, \phi^4$ theory in $ d< 4$). These fixed points define universality classes of phase transitions.

In the context of Wilsonian renormalization and the beta function, stable points are crucial for understanding the behavior of a QFT under RG flow. The stability of a fixed point is determined by the eigenvalues of the stability matrix:
\begin{equation}
M_{ij} = \frac{\partial \beta_i}{\partial g_j}|_{g*}\,,
\end{equation}
where $g_i$ represents the couplings of the theory and $g^*$ is the collection of the critical points. The eigenvalues and eigenvectors of this matrix provide important information about the stability and scaling behavior of the theory near the fixed point.

The eigenvectors of the stability matrix can be determined through the following steps. Let $E$ be an eigenvalue of the stability matrix $M$, and let  $\vec{V}$ be the corresponding vector.  Thus, we have the eigenvalue equation as
\begin{equation}
    M\, \vec{V}  = E\, \vec{V}\,. 
\end{equation}
The eigenvectors, $\vec{V}$, of the stability matrix correspond to specific directions in the space of coupling constants. Each eigenvector is associated with an eigenvalue, which determines the scaling behavior of the corresponding direction under RG flow.
If all eigenvalues are negative, the fixed point is stable (attractive). If some eigenvalues are positive, the fixed point is unstable (repulsive). As a result, at the fixed point: 
\begin{itemize}
    \item A relevant coupling corresponds to a positive eigenvalue $(E>0)$ and corresponding eigenvector describes the relevant direction. Couplings in this direction grow under RG flow away from the fixed point. 
    \item An irrelevant coupling corresponds to a negative eigenvalue $(E<0)$ and corresponding eigenvector describes the irrelevant direction. Couplings in this direction shrink under RG flow and are attracted to the fixed point.
    \item A marginal coupling corresponds to zero eigenvalue $(E=0)$ and corresponding eigenvector describes the marginal direction. The behavior of couplings in this direction requires higher-order analysis to determine stability.
\end{itemize}

Let us now analyze the beta functions of the relativistic Yukawa theory. As can be seen, the only fixed points for $\b$ functions in \eqref{relbeta4} are Gaussian\,. However, when $\l \f^4$ interaction is eliminated\footnote{From now on, we shall call this case as `deformed' relativistic Yukawa theory.}, we obtain the following $\b$ functions:
\begin{equation}
\begin{aligned}
\b_{M^2}&=-2M^2+\frac{g^2 \L^4}{2\pi^2}\frac{\L^2-m^2}{\(\L^2+m^2\)^2}\,, \\
\b_m&=-m +\frac{g^2}{8\pi^2}\frac{\L^4}{\L^2+M^2}\frac{m}{\L^2+m^2}\,, \\
\b_{g}&=\frac{g^3}{8\pi^2}\frac{\L^4}{\L^2+M^2} \frac{\L^2-m^2}{\(\L^2+m^2\)^2}\,.
\end{aligned}
\end{equation}  
It is interesting that the fixed points for the deformed relativistic Yukawa theory are non-Gaussian (apart from the Gaussian points) such that
\begin{equation}
M_*^2=0,\quad m_*=\pm\L\quad \And\quad g_*=\pm4\pi\,. 
\end{equation}

Given our intention to analyze the characteristics of the $\b$ function in the vicinity of these fixed points, we shall proceed with the linearized variational approach. The linearized variations of these parameters can be articulated as  
\begin{equation}
M^2=\delta M^2,\quad m=\pm\L+\delta m,\quad g=\pm4\pi+\delta g\,.
\end{equation}  
Subsequent to the integration of these linearized parameters into the definition of the $\b$ function, we arrive at the resultant expression, manifested in matrix form, as follows:  
\begin{equation}
-b\frac{\partial}{\partial b}\begin{pmatrix}
    \delta M^2\\
    \delta m\\
    \delta g
\end{pmatrix}=\begin{pmatrix}
    -2 & \pm 4\L & 0\\
    \pm \frac{1}{\L} & -1 & \pm\frac{\L}{2\pi}\\
    0 & \pm\frac{4\pi}{\L} &0\\
\end{pmatrix} \begin{pmatrix}
    \delta M^2\\
    \delta m\\
    \delta g
\end{pmatrix}\,.
\end{equation}  
Despite the fact that the constituents of the matrices for each scenario undergo modifications by $\pm$, we determined that each scenario possesses the following eigenvalues:
\begin{equation}
c_1= -3.3553,\quad c_2= 0.177651 - 1.0773 \text{i},\quad c_3=0.177651 + 1.0773 \text{i}\,. 
\end{equation} 
Consequently, only the first eigenvalue is irrelevant whereas the remaining eigenvalues are categorized as relevant and oscillating implying that the UV flow spirals out of the fixed-point region towards the IR. Thus, this system exhibits unstable and oscillatory behavior. From this result, we understand that $\l\f^4$ is the interaction that keeps the system in the Gaussian points.
\subsubsection{Carrollian theory}
A similar analysis can be carried out for the Carrollian regime. However, it is crucial to note that in the Carrollian regime, as seen from propagators and other quantities, fields strictly depend on the energy component but not on the spatial momentum component. Since the fields do not explicitly depend on spatial momentum, any rescaling of the momentum component does not affect them. This implies that we can only scale the energy component. We follow the same notation as in the relativistic regime but decompose it as follows:
\begin{equation}
\begin{aligned}
&w'=b w,\qquad \vec{k'}=\vec{k}\,,\\
&t'=t/b,\hspace{0.8cm}\vec{x'}=\vec{x}\,.
\end{aligned}
\end{equation}
Performing these scalings in the free Carrollian theory yields (setting $\lambda = g= 0$ in \eqref{45}):
\begin{equation}
S_0 =\int d^3 \vec{x'}d t' \[\frac{\f}{2 b}\(-\p_{t'}^2-b^2 M_0^2\)\f+\bar{\P} \(i\slashed{\p}'-b m_0 \)\P\]\,,
\end{equation}
where $\slashed{\p}'=\g^0\del_{t'}$. From this action, we see that, similar to the relativistic case $M^2$ scales as $b^2$ and $m$ scales as $b$\,.  Interestingly, we find that $\f'$=$b^{-1/2}\f $ and $\P'=\P$ which implies that the scaling dimension of  $\f$ ($d_\phi$) is $-1/2$\, while the fermion field $\psi$ does not scale at all (viz $d_\psi = 0$). This serves as our first indication that something intriguing is about to happen. Using this insight, we determine the scaling dimensions of the coupling constants as follows
\begin{equation}
\begin{aligned}
\frac{\l}{4!}\f^4&\rightarrow&\frac{\l}{4!}\f^4 b^{1-4d_\f}&=&b^3 \frac{\l}{4!}\f^4\,,\\
g\f\bar{\P}\P&\rightarrow &g\f\bar{\P}\P b^{1-d_\f-2d_\P}&=&b^{3/2}g\f\bar{\P}\P\,. 
\end{aligned}\label{scalinsCaroll}
\end{equation}
It is interesting that even though the theory 
 contains only marginal couplings in the relativistic case, the coupling constant exhibits a non-trivial dependence on scaling in the Carrollian case. For the sake of completeness, we summarize the scaling behaviors of parameters as \footnote{Note that we neglected the field strength renormalization terms as we did in the relativistic theory. For the details, see Appendix\,. }: 
\begin{equation}
\begin{aligned}
M^2(b)=&b^2[M^2_0+\d_{M^2}]\,,\quad m(b)=b\[m_0+\d_m\]\,, \\
\l(b)=&b^3[\l_0+\d_\l]\,, \hspace{1cm} g(b)=b^{3/2}[g_0+\d_g]\,. \label{Carrollscalings}
\end{aligned}
\end{equation}
\textcolor{red}{}The only remaining step is to determine the $\d$ terms in the Carrollian theory. To proceed with our analysis, we compute the one-loop scalar mass contribution arising from the self-interaction and Yukawa interaction (see diagrams (a) and (b) in Figure \ref{fig:loop}). We adopt the Wick rotation, and since our propagators depend only on $\omega$\,, we set $\omega= i \, \omega_E$: 
\begin{equation}
\begin{aligned}
&-i\d_{M^2}=-(-i g)^2\int \frac{d w\, d^3\vec{k}}{\(2\pi\)^4} \text{Tr}\[\frac{i^2\(\slashed{w}+m\)^2}{\(w^2-m^2\)^2} \]-\frac{i\l}{2}\int \frac{d w\, d^3\vec{k}}{(2\pi)^4}\frac{i}{w^2-M^2}\,,\\
&\text{then,}\hspace{0.5cm}\d_{M^2}=\frac{g^2 K^3}{4\pi^4} \int^{\L}_{\frac{\L}{b}} d w_E \frac{-w_E^2+m^2}{\(w_E^2+m^2\)^2} +\frac{\l K^3}{32\pi^4}\int^{\L}_{\frac{\L}{b}} d w_E\frac{1}{w_E^2+M^2}\,,
\end{aligned}
\end{equation}
where $K^3=\int d^3 \vec{k}$. For simplicity, we again omit the subscript (0) on the parameters\,. As a next step, we analyze the contribution to the fermion mass due to the Yukawa interaction (see diagram (c))
\begin{equation}
\begin{aligned}
&-i\d_m=\(-ig\)^2\int \frac{d w\, d^3\vec{k}}{\(2\pi\)^4}\frac{i}{w^2-M^2}\frac{i m}{w^2-m^2}\,, \\
&\text{then,}\hspace{0.5cm}\d_m=-\frac{g^2 K^3}{16\pi^4}\int^{\L}_{\frac{\L}{b}} d w_E \frac{1}{w_E^2+M^2}\frac{m}{w_E^2+m^2}\,.
\end{aligned}
\end{equation}
Additionally, the contribution to the Yukawa vertex can be expressed as (diagram (d)): 
\begin{equation}
\begin{aligned}
&-i\d_g=(-i g)^3\int \frac{d w\, d^3\vec{k}}{\(2\pi\)^4}\frac{i^3}{w^2-M^2}\frac{w^2+m^2}{\(w^2-m^2\)^2}\,,\\
&\text{then,}\hspace{0.5cm}\d_g= \frac{g^3 K^3}{16\pi^4}\int^{\L}_{\frac{\L}{b}} d w_E  \frac{-w_E^2+m^2}{\(w_E^2+M^2\)\(w_E^2+m^2\)^2}\,. 
\end{aligned}
\end{equation}
As a concluding contribution, we are particularly focused on $\d_\l$ term arising from the self-interaction and fermionic loop (diagrams (e) and (f)):
\begin{equation}
-i\d_{\l}= -6\(-ig\)^4\int \frac{d w\, d^3\vec{k}}{\(2\pi\)^4} \text{Tr}\[\frac{i^4(\slashed{w}+m)^4}{(w^2-m^2)^4}\]+\frac{3}{2}(-i\l)^2\int \frac{d w\,d^3 \vec{k}}{\(2\pi\)^4}\frac{i^2}{\(w^2-M^2\)^2}\,. 
\end{equation}
Consequently, we can interpret the one-loop contribution to $\l$ coupling as
\begin{equation}
\d_\l= \frac{ 3 g^4 K^3}{2\pi^2}\int^{\L}_{\frac{\L}{b}}  dw_E  \frac{w_E^4+m^4-6m^2 w_E^2}{(w_E^2+m^2)^4} -\frac{3\l^2K^3}{32\pi^4}\int^{\L}_{\frac{\L}{b}} d w_E \frac{1}{\(w_E^2+M^2\)^2}\,.
\end{equation}  
Here $\order{\l}\And\order{\l^2}$ terms same as in \cite{Banerjee:2023jpi}, up to some numerical factors. Having established the necessary theoretical framework, we now apply the Wilsonian RG flow methodology that we presented in the previous subsection for relativistic theory to Carrollian theory\,.  Therefore, we are allowed  to derive the $\b$ functions at the 1-loop order:
\begin{equation}
\begin{aligned}
\b_{M^2}&=-2 M^2 -\frac{\l K^3}{32\pi^4}\frac{\L}{\L^2+M^2}+\frac{g^2 K^3}{4\pi^4}\frac{\L\(\L^2-m^2\)}{\(\L^2+m^2\)^2}\,,\\
\b_m&=-m +\frac{g^2 K^3}{16\pi^4}\frac{m \L}{\(\L^2+M^2\)\(\L^2+m^2\)}\,,\\
\b_g&=-\frac{3 g}{2}+ \frac{g^3 K^3}{16\pi^4}\frac{\L\(\L^2-m^2\)}{\(\L^2+M^2\)\(\L^2+m^2\)^2}\,,\\
\b_{\l}&=-3\l +\frac{3\l^2 K^3}{32\pi^4}\frac{\L}{(\L^2+M^2)^2}- \frac{3 g^4 K^3}{2\pi^2} \frac{\L\(\L^4+m^4-6m^2\L^2\)}{(\L^2+m^2)^4}\,. 
\end{aligned}
\end{equation}
Due to the scaling dimensions for Carrollian theory given by \eqref{scalinsCaroll}, there are linear terms in the Carrollian beta functions $\b_g$ and $\b_{\l}$\, which are absent in the relativistic beta functions in \eqref{relbeta4}\,. As in the relativistic regime, our objective is to investigate the implications of the $\l\f^4$ interaction.  To achieve this, we analyze the $\beta$ function by incorporating a four-point interaction term and comparing it with the scenario where this interaction is absent. We begin our analysis with the general case, as previously outlined. When we examine the flow of this theory around Gaussian fixed points, we observe that $M^2\,,m\And g$ are irrelevant\,. However,   the  $\l$ coupling cannot form an eigenvector by itself and instead mixes $M^2$\,ultimately becoming irrelevant as well. The Carrollian regime reveals that the entire structure tends to flow toward the infrared (IR), with no coupling remaining marginal.

By solving for the fixed points of the $\beta$-functions, we find that the  non-Gaussian fixed points  are given by:
\begin{equation}
M_*^2=-\frac{\L^2}{3},\hspace{.2cm} m_*=0,\hspace{.2cm} \l_*=\frac{128\pi^4 \L^3}{9 K^3}, \hspace{.2cm} g_*=0\,.
\end{equation} 
This result aligns with \cite{Banerjee:2023jpi}, up to a numerical factor, which arises from the substitution of $ K=2\pi/a $\,.  To further analyze the behavior of the $\beta$ function near these fixed points, we proceed with a linearized variational approach. The linearized variations  of the parameters are expressed as:    
\begin{equation}
M^2=-\frac{\L^2}{3}+\delta M^2,\hspace{.2cm} m=\delta m,\hspace{.2cm} \l=\frac{128\pi^4 \L^3}{9 K^3}+\delta \l, \hspace{.2cm} g=\delta g\,. 
\end{equation}  
Substituting these linearized parameters into the definition of the $\beta$ function, we obtain the following matrix equation:  
\begin{equation}
-b\frac{\partial}{\partial b}\begin{pmatrix}
    \delta M^2\\
    \delta m\\
    \delta g\\
    \delta \l
\end{pmatrix}=\begin{pmatrix}
    -1 & 0 & 0 & -\frac{3 K^3}{64 \pi^4 \L}\\
    0 & -1 & 0 & 0\\
    0 & 0 & -\frac{3}{2} & 0\\
     -\frac{128\pi^4 \L}{K^3}& 0 & 0 & 3
\end{pmatrix} \begin{pmatrix}
    \delta M^2\\
    \delta m\\
    \delta g\\
    \delta \l
\end{pmatrix}\,. \label{matrixeq}
\end{equation}
The eigenvalues of the resulting matrix are computed as 
\begin{equation}
c_1= -\frac{3}{2},\quad c_2=-1,\quad c_3=1-\sqrt{10},\quad c_4=1+\sqrt{10}\,. \label{eigenvalues}
\end{equation} 
Note that the last two eigenvalues match those found in \cite{Banerjee:2023jpi}. As a result, only the final eigenvalue is relevant, while the remaining eigenvalues are irrelevant. This implies that the fixed points lead to instability. It is interesting to note that the fermionic parameters, namely  $m$ and $g$, become irrelevant in this regime. In this case, neither $M^2$ nor $\l$ can independently form eigenvectors; instead, they couple to each other. Depending on the specific combination of $M^2$  and  $\l$  the couplings can be either relevant or irrelevant.  If one attempts to eliminate the $\lambda \phi^4$ interaction, i.e. obtain the deformed Carrollian Yukawa theory, the $\b$ function provides nothing but Gaussian points. Our findings are summarized in Table \ref{fixed}\,.  

\begin{table}[ht]
\centering
\scalebox{1.}{
\begin{tabular}{|c|c|c|} % Use '|' to add vertical lines between columns
 \hline
 Theory & $ \lambda \neq 0 $ & $\lambda = 0 $ \\ % Header row
 \hline
 Relativistic & Gaussian & Non-Gaussian \\
 \hline
Carrollian &  Non-Gaussian & Gaussian\\
 \hline
\end{tabular}
}
\caption{Fixed points of Yukawa theory\,. } % Caption below the table
\label{fixed} % Optional label for referencing
\end{table}
Interestingly, in the relativistic Yukawa theory, we observed only Gaussian fixed points. This suggests that the system is dominated by free-field behavior, and the interactions are either irrelevant or fine-tuned to zero in the RG flow. However, in the Carrollian regime, these fixed points shift to non-zero values, leading to instability. This may suggest that small fluctuations in coupling constants can drive the system away from equilibrium. Therefore, the Carrollian limit alters the role of this interaction, potentially making it a source of instability rather than stability. Moreover, the instability might also indicate the presence of tachyonic modes in the Carrollian theory. Such behavior might signal the need for additional constraints or mechanisms to stabilize Carrollian field theories, potentially via symmetry breaking or higher-order corrections\,. 
 When we examined the flow of the deformed Carrollian Yukawa theory, we observed the following results: All the couplings are irrelevant which means these Gaussian fixed points are IR. Therefore, we conclude that Carrollian regimes tends to turn marginal couplings to irrelevant couplings and drive them to the IR. Also, we understood that the absence of the $\l$ interaction makes the $M^2$ term irrelevant. Based on these investigations, we also observed that the $\lambda$ interaction alone cannot construct an eigenvector and must couple with the scalar mass in order to flow. However, for specific eigenvalues, it makes the $M^2$ term relevant, as shown in \eqref{matrixeq} and \eqref{eigenvalues}.

Furthermore, when we examined the fixed points in the deformed relativistic Yukawa theory, we found an unstable system which describes that the UV flow spirals out of the fixed-point region towards the IR. These investigations revealed that non-Gaussian fixed points emerge only in the deformed relativistic Yukawa theory for fermion-related parameters, which may suggest a richer structure of interactions and potential phase transitions. The fact that these fixed points vanish in the Carrollian case suggests that interactions may not generate new universality classes, meaning the theory remains perturbatively trivial. This could imply that the deformed Carrollian Yukawa theory does not support a meaningful strongly coupled phase and remains effectively Gaussian at all scales. These conclusions require further study of the deformed version of Carrollian theory.  Consequently, we can infer that the Carrollian limit of any relativistic theory may exhibit behaviors that are fundamentally different from those of the original relativistic theory\,. These results also highlight the challenges of constructing physically meaningful Carrollian field theories and underscores the need for further study to stabilize or modify such theories.

\section{Outlook}
%%%%%%%%%%%%%%%%%%%%%%%%%%%%%%%%%%%%%%%%%%%%%%%%%%
In this paper, we have presented the first example of quantized Carrollian Dirac fermions and investigated their discrete symmetries, including charge conjugation (C), parity (P), and time reversal (T) transformations. We have demonstrated that the Carrollian Yukawa theory provides a natural framework to couple fermions to scalar fields, with the tree-level interactions leading to ultralocal dynamics. By applying the Wilsonian renormalization procedure at one-loop order, we have analyzed the renormalization properties of the theory. Furthermore, our study of the $\beta$ functions reveals important insights into the stability of the fixed points of the theory. These results lay the groundwork for further investigations into the quantum properties of Carrollian field theories, paving the way for exploring more complex interactions and extensions within this framework. It is important to extend the analysis of this project to the following topics:

\begin{itemize}
\item Investigating the quantization of magnetic Carrollian fermions would be an interesting avenue for future research. Since these fermions contain spatial derivatives, their interactions and the Wilsonian renormalization group flow could exhibit significant differences compared to those in the electric theory. A natural starting point for such an analysis would be the classical magnetic fermions studied in \cite{Bergshoeff:2023vfd}\,. 
\item While this paper primarily focuses on massive scalar and fermion fields, an alternative direction would be to consider massless fields and examine the resulting conformal anomaly within the framework of Carrollian Yukawa theory. Understanding the conformal properties of such massless interactions could provide further insights into the role of symmetry and anomalies in Carrollian field theories\,.   
\item A natural extension of the present work involves applying our methods to the Carrollian versions of quantum electrodynamics and quantum chromodynamics. By utilizing the techniques developed in this paper, along with those introduced in \cite{Mehra:2023rmm}\,, one can analyze the structure of gauge interactions in Carrollian settings. At tree level, the ultralocal interactions between fermions and bosonic gauge fields could offer valuable insights into different phases of matter. Additionally, such investigations may shed light on the behavior of quark-gluon plasma (QGP) in extreme conditions, as discussed in the fascinating application to QGP physics presented in \cite{Bagchi:2023ysc}\,.
\item Further exploration of condensed matter applications of quantized fermions within the Carrollian framework is another crucial direction. As demonstrated in \cite{Bagchi:2022eui}, Carrollian dispersion relations appear in various intriguing physical systems, particularly in flat-band physics. These include exotic quantum phases such as spin liquids, fractional quantum Hall systems, and twisted bilayer graphene. Examining quantum corrections from the perspective of Carrollian field theory could provide a novel approach to understanding these systems, potentially leading to new theoretical insights and experimental predictions. Classifying flat-band systems using Carroll symmetry is a novel and interesting avenue for further research. For recent work on this topic, see \cite{Ara:2024vbe} and the references therein. \footnote{We would like to thank Aritra Banerjee for useful comments on flat-band systems.} 
\item Extending this analysis to include supersymmetry represents another significant research direction. A possible starting point would be the supersymmetric field theory formulated in \cite{Koutrolikos:2023evq, Kasikci:2023zdn}\,, which could serve as a foundation for studying the quantum aspects of Carrollian supersymmetric models. Investigating how supersymmetry influences the renormalization and interactions in a Carrollian setting may yield important results for both high-energy physics and mathematical physics.
\item In this work, our focus has primarily been on the continuum limit, as it is directly related to the ultralocal nature of Carrollian field theory. However, it would be valuable to explore the lattice formulation of Carrollian Dirac fields as well. Following the approaches previously developed for scalar fields in \cite{deBoer:2023fnj, Banerjee:2023jpi, Cotler:2024xhb} and for fermions in \cite{Ara:2024vbe}, one could investigate how discretization affects the dynamics of Dirac fermions with Yukawa-like interactions on a lattice and whether lattice-based techniques can provide new insights into condensed matter applications and non-perturbative aspects of Carrollian quantum field theory. An exciting first step in this direction comes from the study of flat band spectra; see \cite{Ara:2024vbe} for details.
\end{itemize}

\begin{acknowledgments}
We would like to thank Stefan Prohazka and Aditya Mehra for their useful comments and feedback. U. Z. is supported by TUBITAK - 3501 - Career Development Program (CAREER) with Grant No. 125F024.
\end{acknowledgments}

%%%%%%%%%%%%%%%%%%%%%%%%%%%%%%%%%%%%%%%%%%%%%%%%%%

%%%%%%%%%%%%%%%%%%%%%%%%%%%%%%%%%%%%%%%%%%%%%%%%%%
\appendix
%%%%%%%%%%%%%%%%%%%%%%%%%%%%%%%%%%%%%%%%%%%%%%%%%%
\section{} \label{Appendix}
%%%%%%%%%%%%%%%%%%%%%%%%%%%%%%%%%%%%%%%%%%%%%%%%%%
In this Appendix, we will provide the detailed  analysis of the beta functions. As discussed in the main text, we have neglected the momenta of the external legs by setting $p=0$\,. However, for a comprehensive analysis, including the field strength renormalization, it is necessary to account for contributions arising from $p\neq 0$\,. Here, we also demonstrate that the relativistic beta functions can be rigorously derived using the Wilsonian RG approach.

To begin, let us consider the expression given in \eqref{loop1rel}\,. This term can be rewritten in a more tractable form by introducing Euclidean momentum and employing Feynman parameters, as follows 
\begin{equation}
\begin{aligned}
&\,\left( p^2\,\delta_\phi -\d_{M^2}\right)= \int \frac{d^4 k_E}{(2\pi)^4}\, \left(- \frac{\lambda}{2(k_E^2 + M^2)}   + 4\, g^2\, \int_{0}^{1}\, dx\, \frac{k_E^2 - \Delta_1}{(k_E^2 + \Delta_1)^2} \right)\,, \label{scalarself1}
\end{aligned}
\end{equation}
where we first use $k' = k - x\,p$\,, then switch to the Euclidean momentum via $k'^0 = i\,k_E^0$\,. Here\,,   $\Delta_1$ is a shorthand notation for  
\begin{equation}
\begin{aligned}
\Delta_1= m^2 + p^2\, x(x-1)\,. 
\end{aligned}
\end{equation}
Expansion about $p^2 = 0$, we obtain the following result
\begin{equation}
\begin{aligned}
&\,\left( p^2\,\delta_\phi -\d_{M^2}\right)= \int^{\L}_{\L/b} \frac{d^4 k_E}{(2\pi)^4}\, \left[\frac{ - \lambda}{2(k_E^2 + M^2)}   + 4\, g^2\, \left( \frac{k_E^2-m^2}{k_E^2 + m^2 }  + \frac{p^2 (3\,k_E^2-m^2)}{6(k_E^2 + m^2)^3}\right)\right]\,. 
\end{aligned}
\end{equation}
Note that since $\L^2\gg p^2$\,, we retain the same integration boundaries\,. As a result, we obtain 
\begin{equation}
\begin{aligned}
\d_{M^2} &= \int^{\L}_{\L/b} \frac{d^4 k_E}{(2\pi)^4}\, \left[\frac{ \lambda}{2(k_E^2 + M^2)}   - 4\, g^2\, \left( \frac{k_E^2-m^2}{k_E^2 + m^2 } \right)\right]\,,  \\
\delta_\phi &= 4\, g^2\, \int^{\L}_{\L/b} \frac{d^4 k_E}{(2\pi)^4}\,\frac{3\,k_E^2 - m^2}{6\, (k_E^2 + m^2)^3}\,. 
\end{aligned}
\end{equation}
The self-energy diagram for the Dirac field \eqref{loop1rel2} can be rewritten as 
\begin{equation}
\begin{aligned}
\left(\slashed{p}\, \delta_\psi - \delta_m\right) = g^2\, \int_0^1\, dx\, \int^{\L}_{\L/b} \frac{d^4 k_E}{\(2\pi\)^4} \frac{x\, \slashed{p} + m}{(k_E^2 + \Delta_2)^2}\,,
\end{aligned}
\end{equation}
where 
\begin{equation}
\begin{aligned}
\Delta_2= m^2 + p^2\, x(x-1) - x (m^2 - M^2)\,. 
\end{aligned}
\end{equation}
Here, we  applied same technique as in \eqref{scalarself1}\,.  Expanding around $p^2 = 0$ yields
\begin{equation}
\begin{aligned}
\delta_\psi &=   g^2\,  \int^{\L}_{\L/b} \frac{d^4 k_E}{\(2\pi\)^4}\, \int_0^1 dx \frac{x}{\(k_E^2 +m^2 - x (m^2 - M^2) \)^2}\,,  \\
\delta_m & = - g^2\int^{\L}_{\L/b} \frac{d^4 k_E}{\(2\pi\)^4}\, \frac{m}{(k_E^2 + m^2)(k_E^2 + M^2)^2}\,. 
\end{aligned}\label{A7}
\end{equation}
Similarly, the vertex terms can be computed as
\begin{equation}
\begin{aligned}
\d_g=& -g^3\int^{\L}_{\L/b} \frac{d^4 k_E}{\(2\pi\)^4} \frac{k_E^2-m^2}{\(k_E^2+M^2\)\(k_E^2+m^2\)^2}\,, \\
\d_\l=& -\frac{3\l^2}{2}\int^{\L}_{\L/b} \frac{d^4 k_E}{\(2\pi\)^4} \frac{1}{\(k_E^2+M^2\)^2} + 24 g^4 \int^{\L}_{\L/b} \frac{d^4 k_E}{\(2\pi\)^4} \frac{k_E^4+m^4-6m^2 k_E^2}{(k_E^2+m^2)^4}\,.
\end{aligned}
\end{equation}
Here, we observe that the $\d_\psi$ term in \eqref{A7} depends non-trivially on $x$. Therefore, we will investigate two cases: the exact results\footnote{First, we complete the $x$ integration and then compute $\b$ functions.} and the limit $\L^2\gg m^2, M^2$ . The exact results yield the following $\b$ functions:

\begin{equation}
\begin{aligned}
\b_{M^2}=&-2 M^2+\frac{g^2\L^4}{2\pi^2}\frac{\L^2-m^2}{\(\L^2+m^2\)^2}-\frac{\l}{16\pi^2}\frac{\L^4}{\L^2+M^2}+\frac{M^2 g^2\L^4}{12\pi^2}\frac{3\L^2-m^2}{\(\L^2+m^2\)^3}\\
\b_m=&-m +\frac{m g^2\,\L^4}{8\pi^2\(\L^2+m^2\)\(\L^2+M^2\)}+\frac{m g^2\L^4}{8\pi^2}\[ \frac{1}{\(m^2-M^2\)\(\L^2+M^2\)}+\frac{\log\(\frac{1+M^2/\L^2}{1+m^2/\L^2}\)}{\(m^2-M^2\)^2} \] \\
\b_g=&\frac{g^3\L^4}{8\pi^2}\frac{\L^2-m^2}{\(\L^2+m^2\)^2\(\L^2+M^2\)}+\frac{g^3\L^4}{24\pi^2}\frac{3\L^2-m^2}{\(\L^2+m^2\)^3}\, \nonumber \\
&+ \frac{g^3\L^4}{8\pi^2}\[ \frac{1}{\(m^2-M^2\)\(\L^2+M^2\)}+\frac{\log\(\frac{1+M^2/\L^2}{1+m^2/\L^2}\)}{\(m^2-M^2\)^2} \]\\
\b_\l=&\frac{3\l^2}{16\pi^2}\frac{\L^4}{\(\L^2+M^2\)^2}-\frac{3 g^4\L^4}{\pi^2}\frac{\L^4-6\L^2 m^2+m^4}{\(\L^2+m^2\)^4}+\frac{\l g^2\L^4}{6\pi^2}\frac{3\L^2-m^2}{\(\L^2+m^2\)^3}
\end{aligned}
\end{equation}
However, there is no analytical solution for the fixed points of these $\b$ functions, neither for the Yukawa theory nor for the deformed Yukawa theory\,. If we consider a  heavy scalar-light fermion interaction regime, where  $m \ll M,\L$, the  new $\b$ functions are given by:
\begin{equation}
\begin{aligned}
\b_{M^2}=&-2 M^2+\frac{g^2\L^2}{2\pi^2}-\frac{\l}{16\pi^2}\frac{\L^4}{\L^2+M^2}+\frac{M^2 g^2}{4\pi^2}\\
\b_m=&-m +\frac{m g^2}{8\pi^2}\frac{\L^2}{\L^2+M^2}+\frac{m g^2\L^4}{8\pi^2}\[ -\frac{1}{M^2\(\L^2+M^2\)}+\frac{\log\(1+M^2/\L^2\)}{M^4} \] \\
\b_g=&\frac{g^3}{8\pi^2}\frac{\L^2}{\L^2+M^2}+\frac{g^3}{8\pi^2}+ \frac{g^3\L^4}{8\pi^2}\[ -\frac{1}{M^2\(\L^2+M^2\)}+\frac{\log\(1+M^2/\L^2\)}{M^4} \]\\
\b_\l=&\frac{3\l^2}{16\pi^2}\frac{\L^4}{\(\L^2+M^2\)^2}-\frac{3 g^4}{\pi^2}+\frac{\l g^2}{2\pi^2}
\end{aligned}
\end{equation}
Unfortunately, even this simplification does not yield an analytical solution for the fixed points of these $\b$ functions. This remains true for the deformed Yukawa theory as well\,. 

As the next case,  we assume $\L^2 \gg m^2, M^2 $\,. Performing  the $x$ integration after taking this limit, yields the following $\b$ functions:
\begin{equation}
\begin{aligned}
\b_{M^2}=&-2 M^2 +\frac{g^2\L^2}{2\pi^2}-\frac{\l \L^2}{16\pi^2}+\frac{M^2 g^2}{4\pi^2}\\
\b_m=&-m +\frac{3m g^2}{16\pi^2}\\
\b_g=&\frac{5 g^3}{16\pi^2} \\
\b_{\l}=& \frac{3\l^2}{16\pi^2}-\frac{3g^4}{\pi^2}+\frac{\l g^2}{2\pi^2}\,.
\end{aligned}
\end{equation}
Note that the last two beta functions coincide with the well-known results in standard QFT texts\,. We may drop the last term in $\b_{M^2}$ due to the condition $M^2\ll\L^2$. Since g corresponds only to Gaussian fixed point, its contribution is irrelevant. Furthermore, as the fixed point for $M^2$ is a Gaussian, our procedure is valid.

Applying the same procedure to the Carrollian regime and proposing $K=a\L$\,, we obtain the following $\b$ functions:
\begin{equation}
\begin{aligned}
\b_{M^2}=&-2 M^2 +\frac{a^3g^2\L^2}{4\pi^4}-\frac{a^3 \l \L^2}{32\pi^4}+\frac{a^3 M^2 g^2}{8\pi^4}\\
\b_m=&-m +\frac{a^3 3m g^2}{32\pi^4}\\
\b_g=&-\frac{3 g}{2}+ \frac{a^3 5 g^3}{32\pi^4} \\
\b_{\l}=&-3\l+\frac{a^3 3\l^2}{32\pi^4}-\frac{a^3 3g^4}{2\pi^4}+\frac{a^3 \l g^2}{4\pi^4}
\end{aligned}
\end{equation}
However, the non-Gaussian fixed points for $M^2$ are $\order{\L^2}$, which invalidates the condition  $M^2\ll\L^2$\,. This could suggest that such fixed points might not be physical. Consequently, we are left with only Gaussian fixed points.

\bibliographystyle{utphys}
\bibliography{ecyd}

\end{document}